\documentclass[11pt]{JHEP}
\usepackage{latexsym}
\usepackage{amssymb,amsfonts}
\usepackage{cite,amstext,amsmath}
\usepackage{amsfonts}
\font\mybb=msbm10 at 12pt
\def\bb#1{\hbox{\mybb#1}}
\def\be{\begin{equation}}
\def\ee{\end{equation}}
\def\bseq{\begin{subequations}}
\def\eseq{\end{subequations}}

\def\bea{\begin{eqnarray}}
\def\eea{\end{eqnarray}}

\renewcommand{\theequation}{\arabic{section}.\arabic{equation}}

\preprint{28/12/2021. V2: 01/02/22.  V3: 19/03/22 \\ 
Published JHEP 03 (2022) 122 \\ V4: three misprints corrected.}
\title{3D supersymmetric nonlinear multiple D$0$-brane action and  4D counterpart of multiple M-wave system
}

\author{
Igor Bandos $^{\dagger\ddagger}$ and Unai D.M. Sarraga $^{\dagger}$
\\ \bigskip
$^{\dagger}$ Department of Physics, University of the Basque Country UPV/EHU,
\\
P.O. Box 644, 48080 Bilbao, Spain
 \\
 $^{\ddagger}$
IKERBASQUE, Basque Foundation for Science, Plaza Euskadi 5,
48009, Bilbao, Spain}

\date{}

\abstract{
Found is the complete nonlinear action of multiple D$0$-brane system (mD$0$) in three dimensional  type II   superspace which is invariant under  rigid $D=3$ ${\cal N}=2$ spacetime supersymmetry and under  local worldline supersymmetry generalizing the $\kappa-$symmetry of single D$0$-brane action. We show that a particular representative of this family of actions can be obtained by dimensional reduction of the action of $D=4$ non-Abelian multiwaves (nAmW), the $D=4$ counterpart of 11D multiple M-wave (mM0) action, that we have also constructed in this paper. This reduction results in an action with is nonlinear  due to the presence of a certain function ${\cal M}({\cal H})$ of the relative motion Hamiltonian ${\cal H}$, the counterpart of which enters the 4D nAmW action linearly.
Curiously, the action possesses double supersymmetry also for an arbitrary function ${\cal M}({\cal H})$. In particular for ${\cal M}= \text{const}$ we find a dynamical system describing the sum of single D$0$ action and the action of 1d dimensional reduction of the $D=3$ ${\cal N}=2$ SYM  coupled to the worldline supergravity induced by the embedding of the center of energy motion into the $D=3$ ${\cal N}=2$  superspace.
 }


\begin{document}

\section{Introduction }

The search for supersymmetric action describing the system of nearly coincident Dirichlet-$p$-branes (D$p$-branes or super-D$p$-branes)  of string theory can be followed for almost 25 years. It is not expected to be given just by the sum of $N$  actions of individual D$p$-branes, which are known from late 90th
\cite{Howe:1996mx,Cederwall:1996pv,Aganagic:1996pe,Cederwall:1996ri,Aganagic:1996nn,Bergshoeff:1996tu,Bandos:1997rq}
\footnote{Actually, the subject of \cite{Howe:1996mx} was the derivation of the equations of motion for super-p-branes, including super-D$p$-branes in the frame of superembedding approach \cite{Bandos:1995zw,Sorokin:1999jx,Bandos:2009xy}. This is based on (and sometimes identified with) the so-called STV (worldline superfield) approach to superparticles and superstrings pioneered in \cite{Sorokin:1988jor} (see \cite{Sorokin:1989jj} for early review) and reaching its highest achievement in \cite{Delduc:1992fk,Sorokin:1993pm,Howe:1994he} where the superfield action for heterotic superstring has been constructed. }. The search is for an effective  action which would describe a system of $N$ D$p$-branes and $N^2$ strings which have endpoints attached to the same or different D$p$-branes. In \cite{Witten:1995im} Witten  argued that such a system should allow for the (gauge fixed) description in terms of
$U(N)$ supersymmetric Yang-Mills (SYM) multiplet.

In weak field limit a single D$p$-brane allows for a gauge fixed  description by the Abelian SYM action, to be precise, by the action of maximally supersymmetric $d=p+1$ SYM. In it the scalar (and spinor) fields of SYM multiplet describe (roughly speaking)  the position of the D$p$-brane worldvolume in $D=10$ dimensional spacetime (superspace), while the gauge field provides the low energy description of fundamental superstring. In the case of $N$ (closely located) D$p$-branes one should also account for the contribution of strings having their ends on different branes. The low energy description of such a string would be also provided by vector field, but with a mass gap increasing with separation between branes. In the limit of $N$ coincident D$p$-branes all the $N^2$ gauge fields become massless. Witten suggested \cite{Witten:1995im} that in this limit of the coincident D$p$-brane also the $U(1)^{N^2}$ gauge symmetry of the system should be enhanced to $U(N)$.
The very low energy (gauge fixed) description of the system of $N$ coincident super-D$p$-branes is thus believed to be provided by the action of the maximally supersymmetric $d=(p+1)$ dimensional $U(N)$ supersymmetric gauge theory.

As far as the effective action of the bosonic D$p$-brane is the Dirac-Born-Infeld action \cite{Leigh:1989jq}, in particular, the Born--Infeld action in the case of spacetime filling D$9$-brane \cite{Fradkin:1985qd}, Tseytlin proposed in this context to study a non-Abelian generalization of the Born-Infeld action with symmetric trace prescription \cite{Tseytlin:1997csa}. The widely accepted candidate for the bosonic part of the action of coincident D$p$-brane is the so-called dielectric brane action proposed by Myers in \cite{Myers:1999ps} motivated by consistency of T-duality rules for the background and D$p$-branes fields. It possess an interesting `dielectric brane effect' which consists in polarization of higher $p$-brane changes in its low $p$-brane version (see also the related study in earlier \cite{Emparan:1997rt}). A similar bosonic action for the system of multiple M-theory gravitons   was proposed and studied in
\cite{Janssen:2002vb,Janssen:2002cf,Janssen:2003ri,Lozano:2005kf}. However, the straightforward supersymmetric generalizations of the actions from \cite{Myers:1999ps} and  \cite{Janssen:2002vb,Janssen:2002cf} were searched for decades and still remain unknown.

A very interesting supersymmetric  `-1 quantization' approach to the problem was developed by Howe, Lindstrom and Wulff in \cite{Howe:2005jz,Howe:2006rv} using the so--called boundary fermions.  The quantization of these allows to reproduces the nonabelian algebra generators from their bilinears, so that  the quantization of the dynamical system from \cite{Howe:2005jz,Howe:2006rv,Howe:2007eb}  should reproduce the desired multiple D$p$-brane  action. However, the complete consistent realization of this step  seems to require the quantization of the complete interacting system of supergravity and super-D$p$-brane.

The  search for a supersymmetric mD$p$ counterpart of the single D$p$-brane actions was the subject of
study in \cite{Sorokin:2001av,Drummond:2002kg,Panda:2003dj,Bandos:2012jz,Bandos:2013swa,Bandos:2013uoa,Bandos:2018ntt}.
In \cite{Drummond:2002kg} a supersymmetrized version of $D=3$ Born-Infeld action was described on the basis of superembedding approach \cite{Bandos:1995zw,Sorokin:1999jx,Akulov:1998bq} and generalized action principle \cite{Bandos:1995dw}. Besides that the supersymmetric and $\kappa-$symmetric actions are known for $d=1$ ($p=0$)  non-Abelian multibrane systems only. These include the action for  the system of ten-dimensional (10D)  multiple $0$-branes \cite{Sorokin:2001av,Panda:2003dj} (see \cite{Bandos:2018ntt} for its review and comparative discussion), the action for 11D  multiple M0 (mM$0$ or multiple M-waves) system
\cite{Bandos:2012jz,Bandos:2013uoa} and a candidate for multiple D$0$-brane (mD$0$) action in $D=10$  \cite{Bandos:2018ntt}
(the present study suggests the existence of its nonlinear generalization).

The reason to discuss  11D  mM$0$ system in the context of present discussion on mD$p$-branes is based on the observation \cite{Bergshoeff:1996tu} that
the dimensional reduction of single M$0$-brane, which is 11D massless superparticle \cite{Bergshoeff:1996tu}, reproduces 10D D$0$-brane action.
Then it was natural to expect that the  10D mD$0$ action can be obtained by dimensional reduction from some 11D action which should be a non-Abelian generalization of the M$0$ action.

Such 11D  action was constructed in  \cite{Bandos:2012jz}, essentially by completing the M$0$ (massless superparticle) action, describing the center of energy motion,  by interacting terms involving matrix fields, the same as should describe the relative motion sector of the 10D mD$0$ system. Namely these latter are the fields of 10D $SU(N)$ SYM multiplet dimensionally reduced to $d=1$ (as, according to \cite{Witten:1995im}, are the fields describing mD$0$ system in the limit of very low energy).
This action was interpreted as describing the dynamical system of multiple interacting M-waves (multiple M$0$-branes or mM$0$), which is similar  to the treatment of the bosonic Myers--type 11D action in \cite{Janssen:2002vb,Janssen:2002cf}.
An evidence of  meaningfulness of this mM$0$ action in the String/M-theory context is its  invariance under two supersymmetries, the 11D spacetime  supersymmetry and the worldline supersymmetry generalizing the $\kappa-$symmetry of massless superparticle (M$0$-brane) action. This latter guarantees that the ground state of the system is 1/2 BPS state, i.e. that it preserves 1/2 of the spacetime supersymmetry and thus saturates the so-called BPS bound (the fact that guarantees its stability).
\footnote{The  mM$0$ action of \cite{Bandos:2012jz} can be also discussed in the context of 11D multiple M$p$-brane systems, for which the allowed values of $p$ is expired by $p=0,2,5$. The progress in understanding of the mM$2$ and mM$5$ cases is even more restricted then of mD$p$-branes. Till 2007 it was not even clear  what should play the role of $d=p+1$ $U(N)$ SYM description of the very low energy limit of mD$p$-brane.
Now it is believed that the infrared fixed points of the system of $N$ coincident M2-branes is  described by   Bagger--Lambert--Gustavsson (BLG) model
\cite{Bagger:2006sk,Gustavsson:2007vu,Bagger:2007jr,Bagger:2012jb} for $N=2$ and by  Aharony--Bergman--Jafferis--Maldacena (ABJM) model \cite{Aharony:2008ug,Bagger:2012jb} for $N\geq 2$. The infrared fixed point of  multiple M$5$-brane system should reproduce  an enigmatic $D=6$ $(2,0)$ superconformal theory which, according to conjecture of  \cite{Douglas:2010iu,Lambert:2010iw}, can be described by $D=5$ SYM model if the nonperturbative sector of this is completely accounted for.}

On the other hand, an attempt to obtain an action for 10D  mD$0$ system from this 11D mM$0$ action, performed in
\cite{Bandos:2018ntt}, was not successful. Instead a candidate mD$0$ action was constructed in \cite{Bandos:2018ntt} directly, by coupling of 1d reduction of 10D SU(N) SYM to the supergravity induced by the embedding of the center of mass worldline in 10D type IIA superspace. The result of present study suggests the hint to resolve this problem with dimensional reduction which is important for the treatment of mM$0$ as a decompactification  limit of type IIA mD$0$ system.

The other problem is that both the 11D mM$0$ action of \cite{Bandos:2012jz,Bandos:2013uoa} and the candidate 10D mD$0$ action of \cite{Bandos:2018ntt} are known for the case of flat target superspace only. To study this latter problem on a toy model,  the $D=3$ counterpart of mM$0$ action was constructed in \cite{Bandos:2013swa}, where it was called non-Abelian multiwave system (nAmW) and  was also generalized for the case of curved AdS superspace. In \cite{Bandos:2018qqo} the quantization of $D=3$ nAmW model was performed, which resulted in a system of relativistic field
equations for the bosonic and fermionic fields on the space with additional non-commutative bosonic and non-anticommutative fermionic coordinates. It was noticed their that the structure of $D=4$ counterpart of 3D nAmW
quantum theory promises to be simpler and more transparent due to the intrinsic complex structure characteristic for the $D=4$ ${\cal N}=1$  superspace (and for the  ${\cal N}=2$ $D=3$ one). So the construction and quantization of 4D nAmW system would provide a better basis for studying the quantization of 11D mM$0$ and 10D mD$0$ models which, in its turn, might shed a new light on the structure and properties of String/M-theory.
The $D=4$ nAmW system can also provide a conventional toy model to approach the problem of curved space generalization and dimensional reduction of 11D mM$0$ system.

Motivated by the above arguments, in this paper we first construct the action for  $D=4$  nAmW system in flat ${\cal N}=1$ superspace, the lower dimensional counterpart of the 11D mM$0$-brane, and show that it is invariant under two supersymmetries: the rigid spacetime (target superspace) $D=4$ ${\cal N}=1$  supersymmetry and the local worldsheet supersymmetry generalizing the $\kappa-$symmetry of the massless $D=4$ ${\cal N}=1$ superparticle
\cite{Siegel:1983hh}\footnote{A bit before \cite{Siegel:1983hh}, the $\kappa-$symmetry was discovered in  massive superparticle model with extended ${\cal N}=2$ supersymmetry \cite{deAzcarraga:1982dhu,deAzcarraga:1982njd}. Its identity as worldline supersymmetry was revealed  in  \cite{Sorokin:1988jor,Sorokin:1989jj}.}.

Then we study its dimensional reduction and find an essentially nonlinear action for $D=3$ ${\cal N}=2$ counterpart of the 10D mD$0$-brane which we, abusing a bit the terminology, call  3D mD$0$-brane. The general form of this 3D mD$0$ action includes an arbitrary function ${\cal M}({\cal H})$ of the
so-called relative motion Hamiltonian ${\cal H}$ which in its turn is constructed from the matrix fields
describing relative motion and interaction of the mD$0$ constituents. The action invariance  under the
$D=3$ ${\cal N}=2$ supersymmetry is manifest. Curiously enough, we find that the invariance under the worldline supersymmetry, which generalizes the $\kappa-$symmetry of a single D$0$-brane \cite{deAzcarraga:1982dhu,deAzcarraga:1982njd}, holds in the case of arbitrary function  ${\cal M}(\cal H)$. In particular, in the case of constant ${\cal M}({\cal H})=m$ we reproduce the 3D counterpart of the candidate mD$0$ action constructed in \cite{Bandos:2018ntt}. This suggests that a more general candidates for 10D mD$0$-brane should also exist in 10D type IIA superspace, and we  will turn to this problem in a forthcoming paper.

The rest of this paper is organized according to the table of content.

We use the mostly minus metric conventions and Weyl spinor indices in $D=4$ which are denoted by dotted and undotted symbols from the beginning of Greek alphabet.
The symbols from the middle of the  Greek alphabet denote the 4-vector indices when they do not carry tilde and  $D=3$ vector indices when are covered by tilde.
The bosonic vector and fermionic spinor coordinates of $D=4$ ${\cal N}=1$ are denoted by
\be\label{ZM=}
Z^M= (x^\mu, \theta^\alpha, \bar{\theta}{}^{\dot{\alpha}})\; , \qquad \mu=0,1,2,3\; , \qquad \alpha=1,2 \; , \qquad \dot{\alpha}=1,2 \; ,
\ee
while
\be
Z^{\tilde{M}}= (x^{\tilde{\mu}}, \theta^\alpha, \bar{\theta}{}^{\alpha})\; , \qquad {\tilde{\mu}}=0,1,2\; , \qquad \alpha=1,2  \qquad
\ee
are coordinates of the $D=3$ ${\cal N}=2$ superspace. The Weyl spinor indices are raised and lowered by Levi-Civita type symbols
\be
\epsilon^{\alpha\beta} = i\sigma_2= \left(\begin{matrix}0 & 1 \cr
                                                                                 -1 &0 \end{matrix}\right)= -\epsilon_{\alpha\beta}=
    \epsilon^{\dot{\alpha}{\dot\beta}}= -\epsilon_{\dot{\alpha}{\dot\beta}}.
\ee
For instance
\be
\theta^\alpha = \epsilon^{\alpha\beta}\theta_\beta \; , \qquad \theta_\alpha = \epsilon_{\alpha\beta}\theta^\beta \; , \qquad
\ee
which also applies to  $D=3$ spinors,
and
\be
\bar{\theta}{}^{\dot\alpha} = \epsilon^{\dot{\alpha}\dot{\beta}}\bar{\theta}{}_{\dot{\beta}} \; , \qquad \bar{\theta}{}_{\dot{\alpha}} = \epsilon_{\dot{\alpha}\dot{\beta}}\bar{\theta}{}^{\dot{\beta}} \; .  \qquad
\ee
We use the following representation for the relativistic Pauli matrices (rPMs)
\bea\label{Pauli=}
\sigma_{\mu\,\alpha\dot{\beta}}= ({\bb I}, \sigma_1, \sigma_2, \sigma_3) = \left(\left(
\begin{matrix} 1 & 0 \cr
0 & 1 \end{matrix}
\right) , \left(
\begin{matrix} 0 & 1 \cr
1 & 0 \end{matrix}
\right), \left(
\begin{matrix} 0 & -i \cr
i & \; 0 \end{matrix}
\right), \left(
\begin{matrix} 1 & \; 0 \cr
0 & -1 \end{matrix}
\right)\right)  \nonumber \\ = \tilde{\sigma}^{\mu\,\dot{\alpha}{\beta}}  :=
\epsilon^{\dot{\alpha}\dot{\beta}}\epsilon^{{\beta}{\alpha}}{\sigma}_{\nu\,{\alpha}\dot{\beta}}\eta^{\nu\mu}
\;  \qquad
\eea
which obey
\bea
(\sigma_{(\mu}\tilde{\sigma}_{\nu )})_\alpha{}^\beta:= \frac 1 2 (\sigma_{\mu}\tilde{\sigma}_{\nu }+\sigma_{\nu}\tilde{\sigma}_{\mu })_\alpha{}^\beta =\eta_{\mu\nu} \delta_\alpha{}^\beta\; , \qquad (\tilde{\sigma}_{(\mu}{\sigma}_{\nu )})^{\dot{\alpha}}{}_{\dot{\beta}}=\eta_{\mu\nu}\delta^{\dot{\alpha}}{}_{\dot{\beta}}\; , \qquad \\
\eta_{\mu\nu} ={\rm diag}(+1,-1,-1,-1)=\eta{}^{\mu\nu} \; , \qquad
\eea
and
\be
\sigma_{\mu \alpha \dot{\alpha}}\tilde{\sigma}{}^{\mu \dot{\beta}{\beta}}= 2\delta_{ \alpha}{}^{\beta} \delta^{\dot{\beta}}{}_{\dot{\alpha}}\; .
\ee

As only one of  the $D=4$ relativistic Pauli matrices  \eqref{Pauli=}, $\sigma_2$, is complex and antisymmetric, we can identify the $D=3$ gamma-matrices, which are known to allow for a real symmetric representation (after rising or lowering indices with the charge conjugation matrix)
  with rPMs carrying indices 0, 1 and 3,
\bea\label{s4d=s3d}
\sigma_{\mu\,\alpha\dot{\beta}}= \left(\sigma_{\tilde{\mu}\,\alpha\dot{\beta}}, \sigma_{2\,\alpha\dot{\beta}}\right) \; , \qquad  \tilde{\sigma}_{\mu}^{\dot{\beta}\alpha}= \left(\tilde{\sigma}_{\tilde{\mu}}^{\dot{\beta}\alpha}, \tilde{\sigma}_{2}^{\dot{\beta}\alpha}\right) \, , \qquad \\
\label{g3=s4}\gamma_{\tilde{\mu}\,\alpha{\beta}}=\sigma_{\tilde{\mu}\,\alpha\dot{\beta}}= ({\bb I}, \sigma_1,  \sigma_3) = \left(\left(
\begin{matrix} 1 & 0 \cr
0 & 1 \end{matrix}
\right) , \left(
\begin{matrix} 0 & 1 \cr
1 & 0 \end{matrix}
\right),  \left(
\begin{matrix} 1 & \; 0 \cr
0 & -1 \end{matrix}
\right)\right) \; ,  \qquad \\
\label{tg3=ts4} \tilde{\gamma}_{\tilde{\mu}}^{{\beta}\alpha}=\tilde{\sigma}_{\tilde{\mu}}^{\dot{\beta}\alpha}= ({\bb I}, -\sigma_1,  -\sigma_3) = \epsilon^{{\alpha}\gamma}\epsilon^{{\beta}\delta}\gamma_{\tilde{\mu}\,\gamma{\delta}}\; .  \qquad
\eea

\section{$D=4$ ${\cal N}=1$ massless superparticle in spinor moving frame (Lorentz harmonic) formulation  }

The $D=4$ nAmW action is a lower dimensional counterpart of the 11D mM$0$ action of \cite{Bandos:2012jz} and in principle can be obtained by dimensional reduction of this latter. However, the complicated structure of 11D spinor frame variables used to write  mM$0$ action  makes easier to construct it from the principles of symmetry, which includes super-Poincar\'e supergroup containing rigid spacetime $D=4$ ${\cal N}=1$ supersymmetry, local worldline supersymmetry generalizing the
$\kappa-$symmetry of the massless $D=4$ ${\cal N}=1$ superparticle and $SU(N)$ gauge symmetry realized on the fields of $d=1$ ${\cal N}=2$ SYM multiplet.

\subsection{Spinor moving frame variables (Lorentz harmonics) in $D=4$}
As the  counterpart of 11D mM$0$, the action of 4D nAmW  can be written in its simplest form  with the use of spinor moving frame variables. In $D=4$ these variables, which were also called Lorentz harmonics \cite{Bandos:1990ji} and can be  identified with  Newman-Penrose diad
\cite{Newman:1961qr} (see \cite{Bandos:1991jh}), are the pair of nonvanishing complex spinors $v_\alpha^\pm= (\bar{v}_{\dot{\alpha}}^\pm)^*$ obeying
\be\label{v-v+=1}
v^{\alpha -}v_\alpha^+=1\; ,  \qquad \bar{v}^{\dot{\alpha} -}\bar{v}_{\dot{\alpha}}^+=1\; .
\ee
This condition implies that the complex matrix composed of the columns $v_\alpha^+$, $v_\alpha^-$ is unimodular and hence belongs to the $SL(2,{\bb C})$ group\footnote{The spinor frame variables are complexification of the
harmonic variables of ${\cal N}=2$ harmonic superspace approach to the off-shell description of the
${\cal N}=2$ SYM, matter and supergravity in terms of unconstrained superfields \cite{Galperin:1984av,Galperin:1987ek,Galperin:2001seg}. Hence the name of Lorentz harmonics used in \cite{Bandos:1990ji}.},
\be\label{VinSL}
\epsilon^{\alpha\beta}v_\alpha^+v_\beta^-=1\qquad \Leftrightarrow \qquad V_\alpha^{(\beta)}:=(v_\alpha^+, v_\alpha^-)\; \in\; SL(2,{\bb C})\; .
\ee

This matrix is called spinor (moving) frame matrix because it can be considered as a kind of square root of a vector frame in the following sense.
From the bilinear of these spinors one can construct two real lightlike vectors and two mutually conjugated complex lightlike vectors
\be\label{u--=v-sv-}
u^=_\mu =    v^-\sigma_\mu \bar{v}^-\; , \qquad u^\#_\mu =     v^+\sigma_\mu \bar{v}^+\; , \qquad u^{-+}_\mu =     v^-\sigma_\mu \bar{v}^+\; , \qquad u^{+-}_\mu =     v^+\sigma_\mu \bar{v}^- =( u^{-+}_\mu )^*\;  \qquad
\ee
which obey
\be
u^=_\mu u^{\mu\#}=2\; ,  \qquad u^{+-}_\mu u^{\mu\,-+}=- 2\; , \qquad  other\; contractions\;=0\; .\qquad
\ee
These vectors  form the light-like tetrade of the Newman-Penrose formalism \cite{Newman:1961qr} and can be collected in the $SO(1,3)$ valued moving frame matrix \footnote{To obtain these relations it is useful to notice that, as it follows from \eqref{VinSL},
\\ ${}\hspace{3cm}  V^{-1\alpha}_{(\beta)}=\epsilon^{\alpha\gamma}V_\gamma^{(\delta)} \epsilon_{(\delta)(\beta)}= (v^{\alpha -},-v^{\alpha +}) $ \\  and that the moving frame matrix is expressed  in terms of spinor frame by the relation  \\ ${}\hspace{3cm} U_\mu^{(a)}= \frac 1 2 \sigma_{\mu\alpha\dot\gamma}V^{-1\dot\gamma}_{(\dot\delta)} \tilde{\sigma}{}^{(a)(\dot\delta)(\beta)}V^{-1\alpha}_{(\beta)}\; .$}
\bea
U_\mu^{(a)} \in SO(1,3)\qquad \Leftrightarrow \qquad U_\mu^{(a)}U^{\mu(b)}=\eta^{ab}= \rm{diag} (+1,-1,-1,-1)\; , \qquad \\  u^\#_\mu = U_\mu^{(0)} - U_\mu^{(3)}\; , \qquad u^=_\mu = U_\mu^{(0)} + U_\mu^{(3)}\; , \qquad u^{\mp\pm}_\mu = -U_\mu^{1} \pm  i U_\mu^{2}\; . \qquad
\eea

It is convenient to write \eqref{u--=v-sv-} in equivalent form by representing vectors by $2\times 2$  matrices which just factorize in terms of our spinor frame variables,
\bea
& u_{\alpha \dot{\beta}}^= = u^=_\mu {\sigma}^\mu_{\alpha\dot{\beta}}= 2v_\alpha^- \bar{v}_{\dot{\beta}}^- \; , \qquad & u_{\alpha \dot{\beta}}^\# = u^\#_\mu {\sigma}^\mu_{\alpha\dot{\beta}}= 2v_\alpha^+ \bar{v}_{\dot{\beta}}^+ \; , \qquad \\ & u_{\alpha \dot{\beta}}^{-+} = u^{-+}_\mu {\sigma}^\mu_{\alpha\dot{\beta}}= 2v_\alpha^- \bar{v}_{\dot{\beta}}^+ \; , \qquad & u_{\alpha \dot{\beta}}^{+-} = u^{+-}_\mu {\sigma}^\mu_{\alpha\dot{\beta}}= 2v_\alpha^+ \bar{v}_{\dot{\beta}}^-= (u_{\beta\dot\alpha}^{-+})^* \; . \qquad
\eea

\subsection{$D=4$ massless superparticle action and its $\kappa-$symmetry}

$D=4$ massless superparticle action can be written in the form \cite{Bandos:1990ji}
\be\label{S0D4=}
S^0_{\rm D=4}= \int_{{\cal W}^1}\rho^{\#} v_\alpha^-\bar{v}_{\dot{\alpha}}^- \Pi^{\alpha\dot{\alpha}} = \int \text{d}\tau \rho^{\#}(\tau) v_\alpha^-(\tau) \bar{v}_{\dot{\alpha}}^-(\tau) \Pi_\tau ^{\alpha\dot{\alpha}}\;  ,
\ee
where
\be\label{Pi=dtPi}
\Pi^{\alpha\dot{\alpha}}=\text{d}\tau \Pi_\tau^{\alpha\dot{\alpha}}\; , \qquad
\Pi_\tau^{\alpha\dot{\alpha}}=\partial_\tau x^{\alpha\dot{\alpha}}(\tau)- 2i \partial_\tau \theta^{\alpha}(\tau)\bar{\theta}^{\dot{\alpha}}(\tau) +
 2i \theta^{\alpha}(\tau)\partial_\tau \bar{\theta}^{\dot{\alpha}}(\tau)\;  \qquad
\ee
is the pull-back to the superparticle worldline of the Volkov-Akulov (VA) 1-form
\bea\label{Pi=}
\Pi^{\alpha\dot{\alpha}}= \text{d}x^{\alpha\dot{\alpha}}- 2i \text{d}\theta^{\alpha}\bar{\theta}^{\dot{\alpha}} +
 2i \theta^{\alpha}\text{d}\bar{\theta}^{\dot{\alpha}} =: \Pi^\mu \tilde{\sigma}_\mu^{\dot{\alpha}\alpha}\; ,
 \\
\Pi^\mu = \text{d}x^{\mu}- i \text{d}\theta\sigma^{\mu}\bar{\theta} +
 \theta\sigma^{\mu}\text{d}\bar{\theta} \; ,
\eea
$x^{\alpha\dot{\alpha}}(\tau)= x^{\mu}(\tau)\tilde{\sigma}_\mu^{\dot{\alpha}\alpha}$ and
$\theta^{\alpha}(\tau)=(\bar{\theta}^{\dot{\alpha}}(\tau))^* $ are bosonic vector and fermionic spinor coordinate functions of proper time $\tau$ which define embedding of the superparticle worldline ${\cal W}^1$ in $D=4$ ${\cal N}=1$ superspace $\Sigma^{(4|4)}$,
\be\label{W1in}
{\cal W}^1 \in \Sigma^{(4|4)}\; : \qquad x^{\mu}=x^{\mu}(\tau)\; ,\qquad \theta^{\alpha}=\theta^{\alpha}(\tau)\; ,\qquad  \bar{\theta}^{\dot{\alpha}}=\bar{\theta}^{\dot{\alpha}}(\tau)\; .
\ee
To simplify the notation, we use the same symbols for  coordinate functions and coordinates (e.g. $x^\mu(\tau)$ and $x^\mu$) as well as for the pull-backs of differential forms to the worldline ${\cal W}^1$  and the forms  on target superspace $\Sigma^{(4|4)}$ (see e.g. \eqref{Pi=dtPi} and \eqref{Pi=}).

The bosonic spinor  fields $v_\alpha^\pm=v_\alpha^\pm (\tau)$ in \eqref{S0D4=} are constrained by
\eqref{v-v+=1} and hence define a spinor moving frame attached to the worldline.
In the light of the above discussion,  we  can  write \eqref{S0D4=} in the following equivalent forms
\be\label{S0D4==}
S^0_{\rm D=4}= \int_{{\cal W}^1}\rho^{\#} \frac 1 2  u_{\alpha\dot{\alpha}}^= \Pi^{\alpha\dot{\alpha}} = \int_{{\cal W}^1}\rho^{\#}  u_{\mu}^= \Pi^{\mu} =\int_{{\cal W}^1} \rho^{\#} E^{=} \; . \qquad
\ee
were, at the last stage, we have introduced one of the pull-backs of the 1-forms
of supervielbein adapted to the embedding of worldline in superspace
\bea\label{E--=}
E^{=}&=&   \Pi^{\mu}u_{\mu}^= =  \frac 1 2  u_{\alpha\dot{\alpha}}^= \Pi^{\alpha\dot{\alpha}} =
 \Pi^{\alpha\dot{\alpha}}v_{\alpha}^-\bar{v}_{\dot{\alpha}}^-
\, , \qquad \\ \label{E++=}
E^{\#}&=&   \Pi^{\mu}u_{\mu}^\# =  \frac 1 2  u_{\alpha\dot{\alpha}}^\# \Pi^{\alpha\dot{\alpha}} =
 \Pi^{\alpha\dot{\alpha}}v_{\alpha}^+\bar{v}_{\dot{\alpha}}^+
\, , \qquad \\ \label{E+-=} E^{-+}&=&   \Pi^{\mu}u_{\mu}^{-+} =  \frac 1 2  u_{\alpha\dot{\alpha}}^{-+} \Pi^{\alpha\dot{\alpha}} =
 \Pi^{\alpha\dot{\alpha}}v_{\alpha}^- \bar{v}_{\dot{\alpha}}^+ = ( E^{-+})^* \, , \qquad
 \\
 \label{E-=}
 E^{\mp}&=& \text{d}\theta^\alpha v_\alpha^\mp \; , \qquad \bar{E}^{\mp}= \text{d}\bar{\theta}{}^{\dot\alpha} \bar{v}{}_{\dot\alpha}^\mp \; , \qquad
\eea
The first two real forms, $E^{=}$ and $E^{\#}$, will be used to write the 4D nAmW action.

The action \eqref{S0D4=} is manifestly invariant under $D=4$ ${\cal N}=1$ spacetime supersymmetry. It also  possesses  the local fermionic  $\kappa-$symmetry
in its irreducible form\footnote{The $\kappa-$symmetry of the Brink-Schwarz form of the massless superparticle action \cite{Siegel:1983hh} is infinitely reducible. Its basic relation can be written in the form  $\delta \theta^\alpha =  \tilde{\kappa}_{\dot{\alpha}} \tilde{\sigma}^{\mu\dot{\alpha}\alpha} P_\mu$
where  $P_\mu$ is light-like on the mass shell,  $P_\mu P^\mu =0$. The irreducible form of the $\kappa-$symmetry \eqref{kappa=sym} can be obtained from this by solving the light-likeness conditions in terms of spinor frame variables, $P_\mu=\rho^\# v^-\sigma_\mu \bar{v}^-$ and identifying
$\kappa^+ = 2  \tilde{\kappa}_{\dot{\alpha}}v^{\dot{\alpha}-} \rho^\#$.
}
\bea\label{kappa=sym}
\delta_\kappa x^{\alpha\dot{\alpha}}&=&  2i \kappa^+ v^{\alpha -}  \bar{\theta}{}^{\dot{\alpha}} + 2i\bar{\kappa}{}^+ \theta^\alpha  \bar{v} {}^{\dot{\alpha} -}  \; , \qquad \nonumber \\ \delta_\kappa\theta^\alpha& =& \kappa^+ v^{\alpha -}\; , \qquad    \delta_\kappa\bar{\theta}{}^{\dot{\alpha}}= \bar{\kappa}{}^+  \bar{v} {}^{\dot{\alpha} -} \; , \qquad \nonumber
\\
&& \delta_\kappa v_\alpha^\mp=0 \; , \qquad \delta_\kappa \rho^\#=0 \; , \qquad
\eea
which can be identified with the worldline supersymmetry \cite{Sorokin:1988jor}.

Notice that the transformation of the bosonic coordinate function in \eqref{kappa=sym} can be written in the form
$\delta_\kappa x^{\alpha\dot{\alpha}}=  2i \delta_\kappa\theta^\alpha \bar{\theta}{}^{\dot{\alpha}} - 2i\theta^\alpha  \delta_\kappa\bar{\theta}{}^{\dot{\alpha}}$ or equivalently
\be\label{idelta}
i_\kappa \Pi^{\alpha\dot{\alpha}}:=  i_{\delta_\kappa} \Pi^{\alpha\dot{\alpha}} := \delta_\kappa x^{\alpha\dot{\alpha}}-  2i \delta_\kappa\theta^\alpha \bar{\theta}{}^{\dot{\alpha}}+ 2i\theta^\alpha  \delta_\kappa\bar{\theta}{}^{\dot{\alpha}} =0\;,
\ee
where $i_\delta $ is defined by $i_\delta \rm d= \delta $. See Appendix \ref{diffF} for the discussion on the use of this and other differential form notation.

\subsection{Cartan forms and other gauge symmetries of the spinor frame action}

For our discussion below it is useful to introduce the $SO$(1,3) Cartan forms and to discuss the 1d supergravity induced by embedding of the superparticle worldline into the target superspace. To this end let us notice that to obtain, starting from \eqref{S0D4=}, the complete set of equations of motion one has to vary also the spinor frame variables constrained by \eqref{v-v+=1}. The simplest way is to define the so-called admissible variations which do not break conditions  \eqref{v-v+=1}. As \eqref{v-v+=1} is one complex conditions imposed on 4 complex variables in
$v_\alpha^\mp$, it is easy to conclude that there are three independent complex  admissible variations. These in their turn can be related to the
Cartan forms
\bea
\Omega^{--}=v^{\alpha -}\text{d}v_\alpha^{-}\; , \qquad \bar{\Omega}{}^{--}=\bar{v}^{\dot{\alpha} -}\text{d}\bar{v}_{\dot{\alpha}}^{-}\; , \qquad \\
\Omega^{++}=v^{\alpha +}\text{d}v_\alpha^{+}\; , \qquad \bar{\Omega}{}^{++}=\bar{v}^{\dot{\alpha} +}\text{d}\bar{v}_{\dot{\alpha}}^{+}\; , \qquad \\
\label{om0=}\omega^{(0)}=v^{\alpha -}\text{d}v_\alpha^{+}\; , \qquad \bar{\omega}{}^{(0)}=\bar{v}^{\dot{\alpha} -}\text{d}\bar{v}_{\dot{\alpha}}^{+}\; , \qquad
\eea
which can be used to write the `admissible' derivatives of spinor frame variables which take into account the conditions \eqref{v-v+=1}:
\bea\label{Dv-=}
\text{d}v_\alpha^-= -\omega^{(0)}v_\alpha^{-}+\Omega^{--}v_\alpha^{+}\qquad \Leftrightarrow \qquad  \text{D}v_\alpha^-:= \text{d}v_\alpha^-+\omega^{(0)}v_\alpha^{-}= \Omega^{--}v_\alpha^{+}\; , \\
\label{Dv+=} \text{d}v_\alpha^+= \omega^{(0)}v_\alpha^{+} - \Omega^{++}v_\alpha^{-}\qquad \Leftrightarrow \qquad  \text{D}v_\alpha^+:= \text{d}v_\alpha^+ -\omega^{(0)}v_\alpha^{+}= -\Omega^{++}v_\alpha^{-}\; ,
\eea
and their c.c. relations.

The expressions for admissible variations
\bea\label{iDv-=}
\delta v_\alpha^-= -i_\delta \omega^{(0)}v_\alpha^{-}+i_\delta \Omega^{--}v_\alpha^{+}\qquad \Leftrightarrow \qquad  i_\delta \text{D}v_\alpha^-:= \delta v_\alpha^-+i_\delta\omega^{(0)}v_\alpha^{-}=i_\delta \Omega^{--}v_\alpha^{+}\; , \\
\label{iDv+=} \delta v_\alpha^+= i_\delta \omega^{(0)}v_\alpha^{+}-i_\delta \Omega^{++}v_\alpha^{-}\qquad \Leftrightarrow \qquad  i_\delta \text{D}v_\alpha^+:= \delta v_\alpha^+ -i_\delta\omega^{(0)}v_\alpha^{+}= -i_\delta\Omega^{++}v_\alpha^{-}\; ,
\eea
can be obtained from \eqref{Dv-=} and  \eqref{Dv+=} by formal contraction with variation symbol $i_\delta$ as described in the Appendix \ref{diffF} (in the case of 1-forms this is essentially the substitution $d\mapsto \delta$).

Eqs. \eqref{Dv-=}, \eqref{Dv+=} also define covariant derivatives D in  which the real and imaginary parts of $\omega^{(0)}$ \eqref{om0=} play the role of connections (gauge fields) for $SO(1,1)$ and $U(1)=SO(2)$ transformations.
These gauge transformations can be parametrized by $i_\delta \omega^{(0)}$ and its c.c. $i_\delta \bar{\omega}{}^{(0)}$ in the
expressions for admissible variations \eqref{iDv-=} and \eqref{iDv+=}. These are the gauge symmetries of the massless superparticle action \eqref{S0D4=} if supplemented by the following $SO(1,1)$ scaling of the Lagrange multiplier $\rho^{\#}$,
\be\label{vr++=SO11}
\delta \rho^{\#}=  (i_\delta \omega^{(0)}+ i_\delta \bar{\omega}{}^{(0)})\rho^{\#} \; .
\ee

One more gauge symmetry of this action is complex two parametric transformation
parametrized by $i_\delta \Omega^{++}$ and by its c.c. $i_\delta \bar{\Omega}{}^{++}$,
\bea
{\bb K}_2\, :\qquad  & \delta_{{K}_2}v_\alpha^-=0\; , \qquad \delta_{{K}_2}v_\alpha^+=-i_\delta \Omega^{++} v_\alpha^- \; , & \qquad \nonumber \\
 & \delta_{{K}_2}\bar{v}_{\dot{\alpha}}^-=0\; , \qquad \delta_{{K}_2}\bar{v}_{\dot{\alpha}}^+=-i_\delta \bar{\Omega}{}^{++} \bar{v}_{\dot{\alpha}}^- \; . & \qquad \nonumber
\eea
In the model invariant under the above described $[SO(1,1)\otimes U(1)]\subset \!\!\!\!\!\!\times {\bb K}_2$ transformations, the variables $v_\alpha^\mp$ can be considered as a kind of homogeneous coordinate for the coset of the $Spin(1,3)= SL(2,{\bb C})$ group isomorphic to the ${\bb S}^2$ sphere \cite{Delduc:1991ir,Galperin:1991gk}
\be
\{v_\alpha^\mp \} \; = \;\frac {SL(2,{\bb C})} {[SO(1,1)\otimes U(1)]\subset \!\!\!\!\!\!\times {\bb K}_2}= {\bb S}^2\; ,
\ee
which can be recognized as the celestial sphere of an observer living in four dimensional spacetime.
The reason to introduce the Lagrange multiplier $\rho^\#$, which has a St\"uckelberg-type transformation \eqref{vr++=SO11} under (i.e. can be gauged to unity by) the $SO(1,1)$ gauge symmetry, is to have the above described clear group-theoretical meaning of the bosonic spinor variables.

\subsection{Equations of motion and induced supergravity on the worldline }

The equations of motion which follow from the action \eqref{S0D4=} include (see  Eqs. \eqref{E--=0}--\eqref{E-=0} in Appendix \ref{diffF} for the complete list of equations)
\be\label{E==0}
  E^==E^{+-}=E^{-+}=0\; , \qquad E^-=\bar{E}^{-}=0\; , \qquad
\ee
so that only the pull-backs of one bosonic and two  fermionic projections of the target superspace supervielbein forms remains nonvanishing on the mass shell. These forms
\be\label{E++=sugra}
E^{\#}= \Pi^{\alpha\dot{\alpha}} v_\alpha^+ \bar{v}{}_{\dot\alpha}^+= \text{d}\tau E_\tau^{\#}\; , \qquad E^{+}= \text{d}\theta^\alpha v_\alpha^+ = \text{d}\tau E_\tau^{+}\; , \qquad\bar{E}^{+}= \text{d}\bar{\theta}{}^{\dot\alpha}\bar{v}{}_{\dot\alpha}^+ = \text{d}\tau \bar{E}_\tau^{+}\; \;  \qquad
\ee
describe 1d extended ${\cal N}=2$ supergravity induced on the worldline by its embedding into the target
$D=4$ ${\cal N}=1$ superspace. This statement is aimed to reflect the fact that under the $\kappa-$symmetry \eqref{kappa=sym}, which can be identified with the worldline supersymmetry,
these forms are transformed by
\be\label{wsSG4D}
\delta_\kappa E^\# =-4iE^+\bar{\kappa}{}^++4i \kappa^+\bar{E}{}^+\; , \qquad \delta_\kappa E^+=\text{D} \kappa^+\; , \qquad \delta_\kappa \bar{E}{}^+=\text{D}\bar{\kappa}{}^+\, . \qquad
\ee
This implies
\be\label{wsSG4D=}
\delta_\kappa E_\tau^\# =-4iE^+\bar{\kappa}{}^+ +4i \kappa^+\bar{E}{}^+\; , \qquad \delta_\kappa E_\tau^+= \text{D}_\tau \kappa^+\; , \qquad \delta_\kappa \bar{E}{}_\tau^+= \text{D}_\tau\bar{\kappa}{}^+\, , \qquad
\ee
which is the characteristic transformation of 1d ${\cal N}=2$ supergravity supermultiplet.

These 1-forms are constructed from the coordinate functions of the superparticle which define the embedding of the worldline into the target superspace \eqref{W1in} and spinor moving frame fields. Hence the name of supergravity induced by  embedding.

The use of induced supergravity allows to couple a one-dimensional supersymmetric matter multiplets  to massless superparticle in a way that preserves its local fermionic $\kappa-$symmetry. Basically, the way consists in coupling of a given multiplet to the above described induced supergravity. The action for 4D nAmW system, the counterpart of 11D mM0 action from \cite{Bandos:2012jz}, is constructed on this way using the 1d reduction of the
3d ${\cal N}=2$ by coupling of $SU$(N) SYM multiplet to the induced supergravity.
We present this action and its worldline supersymmetry in the next section.

\section{$D=4$ ${\cal N}=1$ non-Abelian multiwave action and its worldline supersymmetry  }

The action of 4D non-Abelian multiwave (nAmW) system, which serves as a counterpart of the 11D mM$0$ action, can in principle be obtained from this latter by dimensional reduction. However, as we have already noticed, due to complicated structure of the 11D moving frame variables, it was easier to construct it directly using the procedure described in \cite{Bandos:2013swa} for $D=11$ and in \cite{Bandos:2013uoa} for $D=3$. On this way we found the following 4D nAmW  action
\begin{eqnarray}
\label{SmM0=4D} S^{\rm D=4}_{\rm nAmW} &=& \int\limits_{{\cal W}^1} \rho^{\#}{E}^{=} + {1\over \mu^6} \int\limits_{{\cal W}^1} (\rho^{\#})^3
\text{tr}\left(\bar{\widetilde{{\bb P}}}{\rm D} {\widetilde{\bb Z}} + {\widetilde{{\bb P}}}{\rm D} \bar{\widetilde{\bb Z}} - {i\over 8} {\rm D}{{\widetilde{\Psi}}}\,  \bar{{\widetilde{\Psi}}} + {i\over 8} {{\widetilde{\Psi}}} {\rm D} \bar{{\widetilde{\Psi}}}  \right) + \quad \nonumber \\  &&+ {1\over \mu^6} \int\limits_{{\cal W}^1} (\rho^{\#})^3 \left( {E}^{\#} \widetilde{{\cal H}} + i {E}^{+}
\text{tr}(\bar{{\widetilde{\Psi}}}  \bar{\widetilde{{\bb P}}}+  {\widetilde{\Psi}} [{\widetilde{\bb Z}},  \bar{\widetilde{\bb Z}}])+ i \bar{E}^{+}
\text{tr}({{\widetilde{\Psi}}}  {\widetilde{{\bb P}}}+  \bar{{\widetilde{\Psi}}} [{\widetilde{\bb Z}},  \bar{\widetilde{\bb Z}}]\right) , \qquad
\end{eqnarray}
where
\begin{eqnarray}
\label{HSYM=4D} && \widetilde{{\cal H}}=   \text{tr}\left( {\widetilde{{\bb P}}} \bar{\widetilde{{\bb P}}} +  [{\widetilde{\bb Z}},  \bar{\widetilde{\bb Z}}]^2 -
{i\over 2} {\widetilde{\bb Z}}{ {\widetilde{\Psi}}}{ {\widetilde{\Psi}}} + {i\over 2} \bar{\widetilde{\bb Z}} \bar{{\widetilde{\Psi}}}  \bar{{\widetilde{\Psi}}} \right) \; .
\qquad
\end{eqnarray}

The first term of \eqref{SmM0=4D} coincides with the $D=4$ massless superparticle action \eqref{S0D4=}, \eqref{S0D4==} which can be interpreted now as describing the center of energy movement of the interacting  multiwave system. The remaining part of the nAmW action, proportional to the dimensional parameter  $\frac 1 {\mu^{6}}$ \footnote{Notice that  $\mu$ is of dimension of mass in our physical system of units with  $c=\hbar =1$. Then  $[\mu^{-6}]=M^{-6}=L^6$. See below for comment on the dimension of matrix fields. }  contains bosonic and fermionic matrix fields. These are traceless $N\times N$ (i.e. $[su(N)]^c=sl(N,{\bb C})$ valued) bosonic matrix fields
\begin{eqnarray}
\label{bZ=bZ0++} & {\widetilde{\bb Z}}= \widetilde{\bb Z}_{0|\#}:=\widetilde{\bb Z}_{0|+2}\, , \qquad &  \bar{\widetilde{\bb Z}}= \bar{\widetilde{\bb Z}}_{\#|0}:=
\bar{\widetilde{\bb Z}}_{+2|0}= ({\widetilde{\bb Z}})^\dagger \, , \qquad  \\
\label{bP=bP++3}
& \widetilde{{\bb P}}= \widetilde{{\bb P}}_{+|+3}\, ,\qquad  & \bar{\widetilde{{\bb P}}}= \bar{\widetilde{{\bb P}}}_{+3|+}= (\widetilde{{\bb P}})^\dagger\, ,\qquad  \end{eqnarray}
\footnote{The symbols of the matrix fields of nAmW model are covered by tilde here to distinguish them form the matrix fields of 3D mD$0$ model which are inert under $SO(1,1)$ but carry $U(1)$ charges, see \eqref{bZ0++->bZ}--\eqref{bPsi=bPsi+2+}. One might wonder why we do not make this redefinition in this section, without accompanying it by dimensional reduction. The answer is that such a redefinition will result in appearance of the derivative of the St\"uckelberg fields in the action and we found such form of the action less convenient.} and traceless $N\times N$ fermionic matrix fields
\begin{eqnarray}\label{bPsi=bPsi+2+1} &   { {\widetilde{\Psi}}}=
{\widetilde{\Psi}}_{\# |+}\,  , \qquad & \bar{{ {\widetilde{\Psi}}}}=
\bar{{\widetilde{\Psi}}}_{+|\#}= ({\widetilde{\Psi}} )^\dagger\,  . \qquad
 \end{eqnarray}
Here the sign subindices (equivalent to the opposite sign superindices, e.g. ${\widetilde{\Psi}}_{\# |+}={\widetilde{\Psi}}^{= |-}$) indicate the transformation properties of the matrix fields under the $GL(1,{\bb C})= SO(1,1)\times U(1)$ transformations acting on the spinor frame variables $v_\alpha^\mp=v_\alpha^{(\mp |0)}$ and  $\bar{v}_{\dot\alpha}{}^\mp=\bar{v}_{\dot\alpha}{}^{(0|\mp )}\;$. These latter enter, besides the first term, also in the remaining part of the action  \eqref{SmM0=4D} through the induced supergravity  1-forms:
1d supervielbein $E^\#$, $E^+,\bar{E}^+$ \eqref{E++=sugra} and  Cartan forms $\omega^{(0)}$, $\bar{\omega}{}^{(0)}$ \eqref{om0=} which enter the covariant derivatives
\begin{eqnarray}
\label{DZ:=} & {\rm D} {\widetilde{\bb Z}}=  {\rm d}{\widetilde{\bb Z}} + 2\bar{\omega}{}^{(0)} {\widetilde{\bb Z}} + [\, {\bb A}\, , \, {\widetilde{\bb Z}} \,] \; , \qquad & {\rm D} \bar{\widetilde{\bb Z}}=  {\rm d}\bar{\widetilde{\bb Z}} + 2{\omega}{}^{(0)} \bar{\widetilde{\bb Z}} + [\, {\bb A}\, , \, \bar{\widetilde{\bb Z}} \,] \; , \qquad  \\
\label{DPsi:=}  & {\rm D}
{{\widetilde{\Psi}}}={\rm d}{{\widetilde{\Psi}}}+ (2\omega^{(0)}+\bar{\omega}{}^{(0)}) {{\widetilde{\Psi}}} + [\, {\bb A}\, , \, {{\widetilde{\Psi}}} \,]  \;  ,     \qquad
& {\rm D}
\bar{{\widetilde{\Psi}}}={\rm d}\bar{{\widetilde{\Psi}}}+ (\omega^{(0)}+2\bar{\omega}{}^{(0)}) \bar{{\widetilde{\Psi}}} + [\, {\bb A}\, , \, \bar{{\widetilde{\Psi}}} \,]  \;  ,     \qquad
  \end{eqnarray}
as connection for the  $GL(1,{\bb C})= SO(1,1)\times U(1)$ transformations.

The $SU(N)$ connection 1-form $ {\bb A}={\rm d}\tau {\bb A}_\tau$, with  the traceless antihermitian $N\times N$ matrix field ${\bb A}_\tau$, enters the action \eqref{SmM0=4D} inside the covariant derivatives \eqref{DZ:=},   \eqref{DPsi:=} and their hermitian conjugate.

The dimensions of the matrix matter fields are
\be
{}[ {\widetilde{\bb Z}}]=M =[ \bar{\widetilde{\bb Z}}] \; , \qquad {}[ \bar{\widetilde{\bb P}}]=M^2 =[ {\widetilde{\bb P}}] \; , \qquad [ {{\widetilde{\Psi}}}]=M^{3/2}= [ \bar{{\widetilde{\Psi}}}]\; .
\ee
The choice of such non-canonical dimension allows us to reduce the dependence on the dimensional parameter $\mu$, with dimension of mass $[\mu]=M$,  to an overall multiplier in front of the part  of the action including the matrix fields.

The action (\ref{SmM0=4D}) is invariant under the local worldline supersymmetry transformations which act on the center of energy fields as a bit deformed version of the $\kappa-$symmetry of massless superparticle
\bea\label{w-sh=susyX}
\delta_\epsilon x^{\alpha\dot{\alpha}}&=&  2i \delta_\epsilon\theta^\alpha \bar{\theta}{}^{\dot{\alpha}} - 2i\theta^\alpha  \delta_\epsilon\bar{\theta}{}^{\dot{\alpha}}   + v^{\alpha +} \bar{v}{}^{\dot{\alpha}+ }
  i_\epsilon E^{=} \; , \qquad \nonumber \\ \delta_\epsilon\theta^\alpha& =& \epsilon^+ v^{\alpha -} \; , \qquad    \delta_\epsilon\bar{\theta}{}^{\dot{\alpha}}= \bar{\epsilon}^{+}  \bar{v} {}^{\dot{\alpha} -} \; , \qquad \nonumber
\\
  &&\delta_\epsilon v_\alpha^\mp= 0 \; , \qquad   \delta_\epsilon \bar{v}_{\dot\alpha}^\mp=0 \; , \qquad \delta_\epsilon \rho^\#=0 \; , \qquad
\eea
with
\be\label{ieE--==}
i_\epsilon E^{=} =
  {3i\over 2\mu^6 }  (\rho^{\#})^2   \,
 {\rm tr}\left( \epsilon^{+}\bar{{\widetilde{\Psi}}} \bar{{\widetilde{{\bb P}}}}  +  \bar{\epsilon}{}^{+}{\widetilde{\Psi}} {\widetilde{{\bb P}}} -
 (\epsilon^{+}{\widetilde{\Psi}} + \bar{\epsilon}{}^{+}\bar{{\widetilde{\Psi}}} )[{\widetilde{\bb Z}}, \bar{\widetilde{\bb Z}}]
  \right)\; ,
\ee
and on the matrix fields by
\begin{eqnarray}
\label{susy-Z}  \delta_\epsilon {\widetilde{\bb Z}}  & =& -i \epsilon^{+} \bar{{\widetilde{\Psi}}} \; , \qquad  \delta_\epsilon
\bar{\widetilde{\bb Z}}   =- i \bar{\epsilon}{}^{+} {\widetilde{\Psi}} \; , \qquad
\\ \label{susy-P}
 \delta_\epsilon {\widetilde{{\bb P}}}   &= &  i \epsilon^{+} [{{\widetilde{\Psi}}}, {\widetilde{\bb Z}}]+i  \bar{\epsilon}{}^{+} [\bar{{\widetilde{\Psi}}}, {\widetilde{\bb Z}}]  \; ,\qquad  \delta_\epsilon \bar{\widetilde{{\bb P}}}   = - i \epsilon^{+} [{{\widetilde{\Psi}}}, \bar{\widetilde{\bb Z}}]-i  \bar{\epsilon}{}^{+} [\bar{{\widetilde{\Psi}}}, \bar{\widetilde{\bb Z}}]  \; , \qquad \\
\label{susy-Psi}  \delta_\epsilon {\widetilde{\Psi}} &= &  4\epsilon^{+} \bar{\widetilde{{\bb P}}} + 4 \bar{\epsilon}{}^{+} [{\widetilde{\bb Z}}, \bar{\widetilde{\bb Z}}] \; ,\qquad \delta_\epsilon \bar{{\widetilde{\Psi}}} =  4\bar{\epsilon}{}^{+} {\widetilde{{\bb P}}} + 4 {\epsilon}{}^{+} [{\widetilde{\bb Z}}, \bar{\widetilde{\bb Z}}] \; ,\qquad \\
\label{susy-A}
 \delta_\epsilon {\bb A} & =& i{E}^{\#} (\epsilon^{+}  {\widetilde{\Psi}} -
 \bar{\epsilon}{}^{+}  \bar{{\widetilde{\Psi}}} )   +8i {E}^{+}
 \epsilon^{+}\bar{\widetilde{\bb Z}}  -8i \bar{E}{}^{+}
 \bar{\epsilon}{}^{+}\;    {\widetilde{\bb Z}}
 \; .  \qquad
\end{eqnarray}

An interesting problem is to construct  the Hamiltonian formalism and to perform the quantization of the $D=4$ nAmW system as a preliminary step to approach the problem of quantization of the 11D mM0 system which in its turn might give new insights in the features of String/M-theory. Leaving this for future, below we will study the dimensional reduction of the nAmW action down to $D=3$ which produces a candidate action for the three dimensional counterpart of 10D multiple D$0$-brane, which we call 3D mD$0$-brane.

\section{Nonlinear mD$0$-action in 3D by dimensional reduction of the 4D nAmW  }

\subsection{From $D=4$ massless superparticle  to $D=3$ (counterpart of) super-D$0$-brane}

Now we would like to obtain the action for massive $D=3$ ${\cal N}=2$ superparticle by dimensional reduction of the $D=4$ action \eqref{S0D4=}. To be precise we would like to have the $D=3$ action invariant, besides the extended ${\cal N}=2$ spacetime supersymmetry, under $\kappa-$symmetry and thus being the $D=3$ (counterpart of 10D) super-D$0$-brane.

\subsubsection{Reduction of Brink-Schwarz superparticle }

\medskip

In the case of Brink-Schwarz superparticle action
\be\label{SBS=4D}
S_{\rm BS}^{\rm D=4}= \int \left( p_\mu \Pi^\mu +  \frac 1 2 {\rm d}\tau e p_\mu p^\mu\right)\; ,
\ee
 the dimensional reduction can be performed by using equations of motion for $x^2$ coordinate, which reads d$p_2=0$ and implies that $p_2$ is a constant,
\be\label{p2=m}
{\rm d}p_2=0 \qquad \Rightarrow \qquad p_2=m={\rm const}\; .
\ee

The choice of $x^2$ as a direction of reduction,
\be
x^\mu = (x^{\tilde{\mu}}, x^2)\; , \qquad p_\mu = (p_{\tilde{\mu}}, p_2)\; , \qquad {\tilde{\mu}}=0,1,3 \qquad \leftrightarrow \qquad \mu= 0,1,2,3\; ,
\ee
is convenient because it is in consonance with the representation of  the $D=4$ rPMs \eqref{Pauli=} and its relation with $D=3$ gamma matrices \eqref{g3=s4}.
It corresponds to the split of the 4D VA forms on
\be \label{Pi2=}
\Pi^2={\rm d}x^2-i {\rm d}\theta\sigma^2\bar{\theta}+i\theta\sigma^2{\rm d}\bar{\theta}= {\rm d}x^2-\epsilon_{\alpha\beta} ( {\rm d}\theta^{\alpha} \bar{\theta}^{\beta} -\theta^{\alpha} {\rm d}\bar{\theta}^{\beta})\;
\ee
and 3D VA 1-forms
\be\label{VA=3d}
\Pi^{\tilde{\mu}}={\rm d}x^{\tilde{\mu}}-i {\rm d}\theta{\gamma}^{\tilde{\mu}}\bar{\theta} + i \theta{\gamma}^{\tilde{\mu}}{\rm d}\bar{\theta}=\frac 1 2 \Pi^{\alpha\beta} {\gamma}^{\tilde{\mu}}_{\alpha\beta} \; ,
\ee
which can be represented by symmetric spin-tensor  1-form
\be\label{VA=3d=2x2}
\Pi^{\alpha\beta}= \Pi^{\tilde{\mu}}\tilde{\gamma}_{\tilde{\mu}}^{\alpha\beta}= {\rm d}x^{\alpha\beta}- 2i {\rm d}\theta^{( \alpha}\bar{\theta}{}^{\beta )} + 2i \theta^{( \alpha}{\rm d}\bar{\theta}{}^{\beta )}\; . \qquad
\ee

Substituting the  solution \eqref{p2=m} back into the action \eqref{SBS=4D},
using \eqref{Pi2=} and omitting the total derivative term we find
\be\label{SdAL=}
S_{\rm dAL}^{\rm D=3}= \int \left( p_{\tilde{\mu}} \Pi^{\tilde{\mu}} +  \frac 1 2 {\rm d}\tau e (p_{\tilde{\mu}} p^{\tilde{\mu}}-m^2)\right)+m \int ({\rm d}\theta^\gamma \bar{\theta}_\gamma-\theta^\gamma {\rm d}\bar{\theta}_\gamma )\; .
\ee
This is the $D=3$ counterpart of the De Azc\'arraga-Lukierski ${\cal N}=2$ massive superparticle action
\cite{deAzcarraga:1982dhu,deAzcarraga:1982njd}. This possesses  $\kappa-$symmetry and actually the  $\kappa-$symmetry was first discovered in this example \cite{deAzcarraga:1982dhu}, a bit earlier than for the massless superparticle \cite{Siegel:1983hh}.

In the modern perspective \eqref{SdAL=} is $D=3$ counterpart of super-D$0$-brane action \cite{Bergshoeff:1996tu} or, simplifying terminology, 3D D$0$-brane. The second term in \eqref{SdAL=} is the Wess-Zumino term of the D$0$-brane, the prototype of the Wess--Zumino term of the superstring \cite{Green:1983wt} and of higher super-$p$-branes
(see \cite{Bergshoeff:1987cm,Bergshoeff:1987qx,Achucarro:1987nc},  \cite{Cederwall:1996pv}--\cite{Bandos:1997rq} and refs therein).

\subsubsection{Illuminating   ansatz  for dimensional reduction of spinor moving frame formalism}

Our task now it to perform dimensional reduction of the spinor moving frame action for $D=4$ massless superparticle in such a way that it reproduce 3D D$0$-brane action. (See Appendix B for the warm-up exercise of dimensional reduction reproducing 3D massless superparticle). To this end  it is sufficient to use   the ansatz ({\it cf.} \eqref{v=v*=})
\be\label{v-=v-+}
\begin{cases}
v_\alpha^- ={\rm v}_\alpha^-  - i{\rm v}_\alpha^+  \mu^{=} \; , \qquad \cr v_\alpha^+ ={\rm v}_\alpha^+  \; , \end{cases}\qquad
\begin{cases} \bar{v}_{\dot\alpha}^- ={\rm v}_\alpha^-  + i{\rm v}_\alpha^+  \mu^{=} \; , \qquad \cr \bar{v}_{\dot\alpha}^+ ={\rm v}_\alpha^+ \; , \end{cases} \qquad
\ee
with real spinors ${\rm v}_\alpha^\pm$ obeying
\be\label{v-v+=1=3D}
{\rm v}^{-\alpha}{\rm v}_\alpha^+=1 \; , \qquad ({\rm v}_\alpha^\pm)^*= {\rm v}_\alpha^\pm\; ,
\ee
and some real 1d field $\mu^{=} =\mu^{=} (\tau)= (\mu^{=})^*$.

Let us stress that this ansatz is not suitable for dimensional reduction of the nAmW action due to reasons which we will explain below. However,  we find it quite illuminative, and decided to describe it briefly with hope to  create a feeling of dimensional reduction in spinor moving frame formalism.

Eqs. \eqref{v-=v-+} imply
\be\label{u--=}
u_{\alpha\dot{\beta}}^==2{\rm v}^-_{\alpha}{\rm v}^-_{\beta}+ 2{\rm v}^+_{\alpha}{\rm v}^+_{\beta} (\mu^{=} )^2+ 2i \epsilon_{\alpha\beta}\mu^{=} \quad (\; = (u_{\beta\dot\alpha}^=)^*\; )\;  , \qquad
\ee
and the action \eqref{S0D4==} naturally splits onto the parts containing only 3d VA 1-forms \eqref{VA=3d} and containing
$\Pi^2$  \eqref{Pi2=},
\bea\label{S0D4====}
S^{0}_{\rm D=4}\vert_{\eqref{v-=v-+}}= \frac  1 2 \int \rho^\# ({\rm v}_\alpha^-{\rm v}_\beta^-  +(\mu^=)^2 {\rm v}_\alpha^+{\rm v}_\beta^+)\Pi^{\alpha\beta}-
\int \rho^\# \mu^= \left({\rm d}x^2-\epsilon_{\alpha\beta}({\rm d}\theta^{\alpha}\bar{\theta}^{\beta}-\theta^{\alpha}{\rm d}\bar{\theta}^{\beta})\right)\; . \qquad
\eea
Now let us use the equation of motion for $x^2$ coordinate,
\be\label{rmu=m}
{\rm d}(\rho^\#\mu^=)=0\qquad \Rightarrow\qquad \rho^\#\mu^== m = {\rm const}\; . \ee
Substituting its solution  into  \eqref{S0D4====} we find
\bea\label{S0D3==}
S^{0}_{\rm D=4}\vert_{\eqref{v-=v-+},\eqref{rmu=m}}= \frac  1 2  \int \rho^\# \left({\rm v}_\alpha^-{\rm v}_\beta^-  +\frac {m^2}{(\rho^\#)^2} {\rm v}_\alpha^+{\rm v}_\beta^+\right)\Pi^{\alpha\beta}+
m \int  \epsilon_{\alpha\beta}({\rm d}\theta^{\alpha}\bar{\theta}^{\beta}-\theta^{\alpha}{\rm d}\bar{\theta}^{\beta})\; . \qquad
\eea
The term with $x^2$ has disappeared as it has become the total derivative, and thus we have arrived at the action for $D=3$ superparticle.

The second term in \eqref{S0D3==} is the Wess-Zumino term of the D$0$-brane, the same as in \eqref{SdAL=}.
Notice that $\rho^\#$ can be just removed from the first, kinetic term if we   redefine the 3D spinor moving frame variables, by
\be
{\rm v}_\alpha^2 =\sqrt{\frac {\rho^\#}{m} } {\rm v}_\alpha^- \; , \qquad {\rm v}_\alpha^1 = \sqrt{\frac m {\rho^\#} } {\rm v}_\alpha^+ . \qquad
\ee
The redefined spinors still obey
\be\label{v2v1=1}
{\rm v}^{\alpha 2}{\rm v}_\alpha^1=1\qquad \Longleftrightarrow \qquad {\rm v}^{\alpha p}{\rm v}_\alpha^q =-\epsilon^{pq}\; ,\qquad q=1,2\;
\ee
and hence form the $SL(2,{\bb R})$ valued matrix
\be\label{vq=inSL}
{\rm v}_\alpha^q = ({\rm v}_\alpha^1,{\rm v}_\alpha^2)\;\in\; SL(2,{\bb R}) \; .
\ee

Thus the dimensional reduction of the 4D massless superparticle action \eqref{S0D4==} results in
\bea\label{SD0D3=}
S^{ \rm D0}_{\rm D=3} =: \int {\cal L}^{0}_{\rm D=3}&=& \frac 1 2  m\int {\rm v}^q_{\alpha}{\rm v}^q_{\beta}\Pi^{\alpha\beta}+
m \int  \epsilon_{\alpha\beta}({\rm d}\theta^{\alpha}\bar{\theta}^{\beta}-\theta^{\alpha}{\rm d}\bar{\theta}^{\beta})\;  \qquad \nonumber \\ &=& \frac 1 2 m\int {\rm u}^0_{\alpha\beta}\Pi^{\alpha\beta}+
m \int  \epsilon_{\alpha\beta}({\rm d}\theta^{\alpha}\bar{\theta}^{\beta}-\theta^{\alpha}{\rm d}\bar{\theta}^{\beta})\;  \qquad
\eea
which is the  $D=3$ D$0$-brane action in spinor moving frame formulation \cite{Bandos:2000tg,Bandos:2018ntt}.
In the second line  of \eqref{SD0D3=} we have introduced the matrix
${\rm u}^0_{\alpha\beta}={\rm v}^q_{\alpha}{\rm v}^q_{\beta}$ representing timelike normalized vector from the 3d moving frame attached to the worldline,
\bea\label{u0=}
{\rm u}^0_{\alpha\beta}={\rm u}^{0\tilde{\mu}}{\gamma}_{\tilde{\mu}\alpha\beta}={\rm v}_\alpha^1{\rm v}_\beta^1+{\rm v}_\alpha^2{\rm v}_\beta^2\; , \qquad \\  \label{u12=}
{\rm u}^1_{\alpha\beta}={\rm u}^{1\tilde{\mu}}{\gamma}_{\tilde{\mu}\alpha\beta}=2{\rm v}_{(\alpha}^1{\rm v}_{\beta)}^2\; , \qquad {\rm u}^2_{\alpha\beta}={\rm u}^{2\tilde{\mu}}{\gamma}_{\tilde{\mu}\alpha\beta}={\rm v}_\alpha^1{\rm v}_\beta^1-{\rm v}_\alpha^2{\rm v}_\beta^2
\; . \qquad
\eea
These vectors obey the orthogonality and normalization conditions
\be
{\rm u}^0_{\tilde{\mu}}{\rm u}^{0\tilde{\mu}}=1 \; , \qquad {\rm u}^0_{\tilde{\mu}}{\rm u}^{I\tilde{\mu}}=0 \; , \qquad {\rm u}^I_{\tilde{\mu}}{\rm u}^{J\tilde{\mu}}=-\delta^{IJ} \; , \qquad I,J=1,2\;
\ee
which imply that they form $SO(1,2)$ valued 3D moving frame matrix
\be
{\rm u}_{\tilde{\mu}}{}^{\tilde{a}}=\left({\rm u}_{\tilde{\mu}}{}^{0},{\rm u}_{\tilde{\mu}}{}^{I}\right)\; \in\; SO(1,2)\; . \ee

\subsubsection{Irreducible $\kappa-$symmetry of 3D super-D$0$-brane }

If we take the exterior derivative  (see Appendix A) of the Lagrangian form of the action \eqref{SD0D3=} and do not concentrate on the derivatives of the spinor moving frame variables, we find
\bea\label{dL0D3=}
{\rm d} {\cal L}^{0}_{\rm D=3} =- 2im ({\cal E}^1+i{\cal E}^2) \wedge  (\bar{{\cal E}}^1-i\bar{{\cal E}}^2) + \frac 1 2  m\Pi^{\alpha\beta}\wedge {\rm d}{\rm u}^0_{\alpha\beta}\; ,  \qquad
\eea
where we have used the fermionic part of the pull-back of 3D supervielbein adapted to the embedding
\be\label{cEq=}
{\cal E}^{q}= ({\cal E}^{1},{\cal E}^{2} )={\rm d}\theta^\alpha {\rm v}_\alpha^q \; , \qquad \bar{{\cal E}}{}^{q}= (\bar{{\cal E}}{}^{1},\bar{{\cal E}}{}^{2} )={\rm d}\bar{\theta}^\alpha {\rm v}_\alpha^q \; . \qquad
\ee
Below we will also need one of three bosonic 1-forms of this supervielbein
\bea\label{rmE0=}
{\rm E}^0 = \Pi^{\tilde{\mu}} {\rm u}_{\tilde{\mu}}^0= \frac 1 2 \Pi^{\alpha\beta} {\rm u}_{\alpha\beta}^0 = \frac 1 2 \Pi^{\alpha\beta}({\rm v}_\alpha^1{\rm v}_\beta^1 + {\rm v}_\alpha^2{\rm v}_\beta^2)\; , \qquad {\rm E}^I = \Pi^{\tilde{\mu}} {\rm u}_{\tilde{\mu}}^I= ( {\rm E}^1, {\rm E}^2)\; . \qquad
\eea

Using the formalism described in Appendix \ref{diffF}  we can conclude from \eqref{dL0D3=}  that the action  \eqref{SD0D3=} is invariant under $\kappa-$symmetry defined by
\bea\label{kappa=D0}
i_\kappa {\cal E}^1 = \kappa \; , \qquad i_\kappa {\cal E}^2 =  i\kappa \; , \qquad i_\kappa \bar{{\cal E}}^1 = \bar{\kappa} \; , \qquad i_\kappa \bar{{\cal E}}^2 = -i\bar{\kappa} \; , \qquad \nonumber \\
i_\kappa \Pi^{\alpha\beta}=0\; , \qquad \delta_\kappa {\rm  v}^q_\alpha =0\qquad (\Rightarrow \qquad \delta_\kappa {\rm  u}^0_{\alpha\beta}=0)\; , \qquad
\eea
or, equivalently,
\bea
\delta_\kappa \theta^\alpha = \kappa ({\rm  v}^{\alpha 2} -i {\rm  v}^{\alpha 1}) \; , \qquad \delta_\kappa \bar{\theta}{}^\alpha = \bar{\kappa } ({\rm  v}^{\alpha 2} +i {\rm  v}^{\alpha 1}) \; , \qquad \nonumber \\
\delta_\kappa x^{\alpha\beta} = 2i (\delta_\kappa \theta^{(\alpha}  \,  \bar{\theta}{}{}^{\beta )} - \theta^{(\alpha} \delta_\kappa \bar{\theta}{}^{\beta )})\; , \qquad \delta_\kappa {\rm  v}^q_\alpha =0\; . \qquad
\eea

Let us also observe that the action \eqref{SD0D3=} is invariant under the local $SO(2)$ rotation of the spinor frame variables \eqref{vq=inSL}, which also produce $SO(2)$ rotations mixing the spacelike vectors ${\rm u}_{\tilde{\mu}}{}^{I}=\left({\rm u}_{\tilde{\mu}}{}^{1},{\rm u}_{\tilde{\mu}}{}^{2}\right)$.
Using this $SO(2)$ as an identification relation on the set of spinor moving frame fields ${\rm v}_\alpha^q$,
we can consider these as  homogeneous coordinates of the coset
\be\label{coset=3D}
\frac {SL(2,{\bb R})} {SO(2)} \simeq \frac {SO(1,2)} {SO(2)} \; .
\ee
This is suitable for the description of the massive particle as  $SO(2)$ is the small group of 3D timelike  momentum.

Another observation is that the above $SO(2)$ symmetry is not seen in the original reduction ansatz
\eqref{v-=v-+} for 4D spinor frame variables. It appears in the final action as a kind of emergent symmetry. However, if we try to use this ansatz for dimensional reduction of a more complicated nAmW system,
such a symmetry will not emerge in the final answer. This would imply that the spinor frame variables in this 3D action parametrize the $SL(2,{\bb R})$ group rather then coset \eqref{coset=3D} and hence carry one extra degree of freedom which actually is not unwanted.
This is why for dimensional reduction of nAmW  system we will use a bit more complicated ansatz which we will first describe on a simpler example of the reduction of massless superparticle to $D=3$ super-D$0$-brane.

\subsection{Revising the 4D M$0$ reduction to 3D D0. Properties of 3D spinor frame}

The above described dimensional reduction was based on the ansatz \eqref{v-=v-+} for spinor moving frame variables
which breaks explicitly the $U(1)$ subgroup of the $GL(1,{\bb C})$ gauge symmetry of the 4D spinor moving frame formalism. Furthermore with it, an important $SO(2)$ gauge symmetry of the $D=3$ Lorentz harmonic approach was restored only at final stage as an emergent symmetry of the single super-D$0$-brane action \eqref{SD0D3=}. The origin of this emergent gauge symmetry can be followed to the fact that the original $D=4$ massless superparticle action  involves only one of four moving frame vectors, ${u}_{\mu}^=$; and this would not happen if we apply that ansatz to the reduction of nAmW action \eqref{SmM0=4D} which also involves $u_\mu^\#$.

\subsubsection{$U(1)=SO(2)$ invariant reduction of $D=4$ spinor frame formalism}

In this section we describe a more complicated $SO(2)\simeq U(1)$ invariant ansatz for the reduction of the $D=4$ spinor moving  frame formalism down to $D=3$ which is characterized by the following expressions for reduced  light-like vectors of the 4D moving frame
\bea\label{ru--=}
\rho^\# u_{\alpha\dot\beta}^= =  {\cal M} {\rm u}^0_{\alpha\beta} + i {\cal M}\epsilon_{\alpha\beta}\; , \qquad \\
\label{ru++=}
\frac 1 {\rho^\#  }u_{\alpha\dot\beta}^\#  =  {\cal M} {\rm u}^0_{\alpha\beta} - i {\cal M}\epsilon_{\alpha\beta}\; , \qquad
\eea
where ${\cal M}={\cal M}(\tau)$ is a new real field on the worldline.
As only these moving frame  vectors are present explicitly in the nAmW action, and their reduction does not contain the vectors ${\rm u}_{\alpha\beta}^I =({\rm u}_{\alpha\beta}^1, {\rm u}_{\alpha\beta}^2)$ we can expect that the reduced action will be invariant under the $SO(2)$ rotation mixing these vectors.

The explicit form of the $SO(2)\simeq U(1)$ invariant ansatz expressing  the $D=4$ spinor moving frame variables \eqref{VinSL} in terms of 3D Lorentz harmonics ${\rm v}_\alpha^{q}=({\rm v}_\alpha^{1},{\rm v}_\alpha^{2})$ \eqref{vq=inSL}, which produces \eqref{ru--=} and  \eqref{ru++=} reads
\bea\label{rv-=v2-iv1}
& \sqrt{\rho^\#} v_\alpha^- =\frac {\sqrt{{\cal M}}}{\sqrt{2}}\, \left({\rm v}_\alpha^2 -i {\rm v}_\alpha^1 \right)\; , \qquad
 \frac {1}{\sqrt{\rho^\#}}
v_\alpha^+ =\frac 1{\sqrt{2}\sqrt{{\cal M}}}\, \left({\rm v}_\alpha^1 -i {\rm v}_\alpha^2 \right)\; , \qquad \nonumber
\\
 & \sqrt{\rho^\#} \bar{v}_{\dot\alpha}^- =
\frac {\sqrt{{\cal M}}} {\sqrt{2}}\, \left({\rm v}_\alpha^2 +i {\rm v}_\alpha^1 \right)\; , \qquad
 \frac {1} {\sqrt{\rho^\#}}  \bar{v}_{\dot\alpha}^+ =  \frac {1}{\sqrt{2}\sqrt{{\cal M}}}\, \left({\rm v}_\alpha^1 +i {\rm v}_\alpha^2 \right)\; .  \qquad
\eea
Notice that this implies
\be
\label{bv-=iv+}  \sqrt{\rho^\#} \bar{v}_{\dot\alpha}^- = i \frac {{\cal M}} {\sqrt{\rho^\#}} v_\alpha^+ \; , \qquad
 \frac {1} {\sqrt{\rho^\#}}  \bar{v}_{\dot\alpha}^+ =  i\frac {\sqrt{\rho^\#}} {{\cal M}}v_\alpha^-\;   \qquad
\ee
which reflects the $SO(2) = U(1)$ invariance of the ansatz \eqref{rv-=v2-iv1}.

Actually, \eqref{bv-=iv+} shows that the complete $GL(2,{\bb C})$ gauge symmetry of the 4D Lorentz harmonic formalism is preserved by the ansatz \eqref{rv-=v2-iv1}. Its $SO(1,1)$ subgroup leaves invariant the l.h.s.-s and does not act on the r.h.s.-s, while $U(1) \subset GL(2,{\bb C})$ transformations of the l.h.s. produce the $U(1)=SO(2)$ transformations of 3D spinor frame variables in the r.h.s.

Using \eqref{rv-=v2-iv1}, the  (pull-backs of)  relevant  4D supervielbein forms \eqref{E--=}--\eqref{E-=} can be expressed in terms of  (pull-backs of)
3D supervielbein forms \eqref{rmE0=}, \eqref{cEq=}:
\bea\label{E--=cME0+}
\rho^\# E^= =  {\cal M} {\rm E}^0 + \frac i 2 {\cal M}\epsilon_{\alpha\beta} \Pi^{\alpha\dot{\beta}} &=&  {\cal M} {\rm E}^0 -{\cal M}\Pi^{2}\nonumber \\ &=& {\cal M} {\rm E}^0 - {\cal M}({\rm d}x^2- {\rm d}\theta^\gamma \bar{\theta}_\gamma+ \theta^\gamma {\rm d}\bar{\theta}_\gamma )\; ,   \qquad \\
\label{E++=cME0-}
\frac 1 {\rho^\#} E^\# =  \frac 1 {{\cal M}} {\rm E}^0 - \frac i 2  \frac 1 {{\cal M}} \epsilon_{\alpha\beta} \Pi^{\alpha\dot{\beta}} &=&   \frac 1 {{\cal M}}  {\rm E}^0+ \frac 1 {{\cal M}} \Pi^{2}\nonumber \\ &=&  \frac 1 {{\cal M}} {\rm E}^0 +   \frac 1 {{\cal M}} ({\rm d}x^2 - {\rm d}\theta^\gamma \bar{\theta}_\gamma+ \theta^\gamma {\rm d}\bar{\theta}_\gamma )\; , \qquad \\
\label{E+=cME0-}
\frac 1 {\sqrt{\rho^\#}} E^+ =  \frac 1{\sqrt{2}\sqrt{{\cal M}}}\, \left({\cal E}^1 -i {\cal E}^2 \right) &,& \qquad
\frac 1 {\sqrt{\rho^\#}} \bar{E}^+ =  \frac 1{\sqrt{2}\sqrt{{\cal M}}}\, \left(\bar{{\cal E}}^1 +i \bar{{\cal E}}^2 \right) \; , \qquad
\eea

The dimensional reduction of the massless superparticle action requires to use only the first of these expressions, \eqref{E--=cME0+}. Substituting it into the action and using the equations of motion for $x^2$,
\be
{\rm d}{\cal M}=0\qquad\Rightarrow \qquad {\cal M}=m={\rm const}\;
\ee
we arrive at the  3D super-D$0$-brane action \eqref{SD0D3=}.

\subsubsection{Cartan forms and covariant derivatives in 3D }

To describe the dimensional reduction of the nAmW action we have to introduce the $SL(2,{\bb R})/SO(2)$ Cartan forms
\be
 f^{pq}:= {\rm v}^{\alpha p} {\rm d}{\rm v}_\alpha^q =+ f^{qp}\; , \qquad {\rm v}^{\alpha p}=\epsilon^{pq}{\rm v}^\alpha_q=\epsilon^{\alpha\beta}{\rm v}_\beta^p\; . \qquad
\ee
Due to \eqref{v2v1=1} their matrix is symmetric, $f^{pq}=f^{qp}$ and the derivative of the 3D spinor frame variables are expressed in terms of these by
\be
{\rm d}{\rm v}_\alpha^q= {\rm v}_{\alpha p}f^{pq} \, ,  \qquad
\ee
where ${\rm v}_{\alpha p}=\epsilon_{pq}{\rm v}_{\alpha}^q$, or in more details
\be
{\rm d}{\rm v}_\alpha^1= {\rm v}_\alpha^1f^{21}- {\rm v}_\alpha^2f^{11}\; , \qquad {\rm d}{\rm v}_\alpha^2= {\rm v}_\alpha^1f^{22}- {\rm v}_\alpha^2f^{12}\; . \qquad
\ee

It is important  that $f^{qq}$ transforms as $SO(2)$ connection, while $f^{12}=f^{21}$ and $f^{11}-f^{22}$ forms are covariant under $SO(2)$.
This can be easily seen using the 3d vector frame \eqref{u0=}, \eqref{u12=}
which allows to identify the $SO(2)$ connection
\be\label{u1du2}
{\rm u}^1{\rm d}{\rm u}^2=-{\rm u}^2{\rm d}{\rm u}^1= \frac 1 2 {\rm u}^{1\alpha\beta}{\rm d}{\rm u}^2_{\alpha\beta}=f^{qq}\qquad \Leftrightarrow \qquad {\rm u}^I{\rm d}{\rm u}^J=\epsilon^{IJ}f^{qq} \; ,\qquad \ee
and the covariant forms $f^I=(f^1,f^2)$ forming the vielbein of the  $SO(1,2)/SO(2)$ coset
\be\label{u0duI} f^1:=  {\rm u}^0{\rm d}{\rm u}^1 =-f^{11}+f^{22}\; ,\qquad f^2:= {\rm u}^0{\rm d}{\rm u}^2= 2f^{12}\; . \qquad
\ee

For our discussion in the next section it is important that \eqref{rv-=v2-iv1} and  \eqref{om0=} imply
the following expressions for  4D  connections
\be\label{om0=fqq+dM}
\omega^{(0)} - \bar{\omega}^{(0)} =i{\rm v}^{\alpha q}{\rm d}{\rm v}_\alpha^q=: i f^{qq}\; , \qquad \omega^{(0)} + \bar{\omega}^{(0)} = \frac {{\rm d}\rho^\#}  {\rho^\#}\, - \frac {{\rm d}{\cal M}}  {{\cal M}}\, . \ee
The second of these equations implies  that the  covariant derivative of the St\"ukelberg field of the 4D nAmW system is expressed in terms of derivative of the  field ${\cal M}$:
\be\label{Dr=dcM}
{\rm D}\rho^\#= {\rm d}\rho^\#- (\omega^{(0)} + \bar{\omega}^{(0)}) \rho^\#= \rho^\#  \frac {{\rm d}{\cal M}}  {{\cal M}}\, .  \qquad
\ee

\subsection{${\cal N}=2$ mD$0$-brane action from dimensional reduction of the 4D nAmW}

Let us turn to the problem of the dimensional reduction of the $D=4$ nAmW action \eqref{SmM0=4D}
This procedure simplifies if, before  substituting the above discussed ansatz \eqref{rv-=v2-iv1} for 4D spinor frame variables, we redefine the matrix fields passing to the  $SO(1,1)$ invariant ones by multiplication on  suitable powers of the $D=4$ St\"uckelberg field $\rho^\#$:
\begin{eqnarray}
\label{bZ0++->bZ} & \widetilde{{\bb Z}}= \dfrac{1}{\rho^\#} {\bb  Z} \, , \qquad &  \bar{\widetilde{{\bb Z}}}= \frac 1 {\rho^\#} \bar{\bb  Z} \, , \qquad  \\
\label{bP=bP++3+}
& \widetilde{{\bb  P}}= \dfrac{1}{(\rho^\#)^2} {\bb  P}\, , \qquad   & \bar{\widetilde{{\bb  P}}}= \frac 1 {(\rho^\#)^2} \bar{\bb  P} \, , \qquad    \\
\label{bPsi=bPsi+2+} &  \widetilde{\Psi}=\dfrac{1}{(\rho^\#)^{\frac 32}} {\Psi} \, , \qquad & \bar{\widetilde{ \Psi}}=\frac 1 {(\rho^\#)^{\frac 32}} \bar{{ \Psi}} \, , \qquad
 \end{eqnarray}
Using \eqref{bZ0++->bZ}--\eqref{bPsi=bPsi+2+} one can check that the left and right `charges' of the new matrix fields are opposite,
 \begin{eqnarray}
\label{Z=Z-+} &  {\bb  Z} =  {\bb Z}_{-|+}\, , \qquad &   \bar{\bb  Z} = \bar{\bb  Z}_{+|-}\, , \qquad  \\
\label{P=P-+}
& {\bb  P} =   {\bb  P}_{-|+}\, , \qquad   &  \bar{\bb  P} =  \bar{\bb  P}_{+|-}\, , \qquad    \\
\label{Psi=Psi+-} & {\Psi} =  {\Psi}_{\frac 12|-\frac 12}\, , \qquad &  \bar{{ \Psi}} =  \bar{{ \Psi}}_{-\frac 12|\frac 12}\, , \qquad
 \end{eqnarray}
 which is tantamount to the statement of their $SO(1,1)$ invariance.

The above redefinition clearly produces the terms proportional to d$\rho^\#$ in the action. However, after it is accompanied by the reduction of the $D=4$ spinor moving frame sector with the ansatz \eqref{rv-=v2-iv1}, such derivatives becomes replaced by the derivatives of the new field ${\cal M}$ inert under both $SO(1,1)$ and $SO(2)$ transformations. Indeed, one can easily check, using \eqref{Dr=dcM}, that the covariant derivatives of old and new matrix fields are related by
\bea\label{DZ=4D-3d}
{\rm D}\widetilde{{\bb Z}}=  \frac 1 {\rho^\#}\left({\rm D}{\bb Z}-  \frac {{\rm d}{\cal M}} {{\cal M}} {\bb Z} \right)\; , \qquad \\ \label{DbZ=4D-3d}
{\rm D}\bar{\widetilde{{\bb Z}}}=  \frac 1 {\rho^\#}\left({\rm D}\bar{{\bb Z}}-  \frac {{\rm d}{\cal M}} {{\cal M}} \bar{{\bb Z}} \right)\; , \qquad \\
\label{DPsi=4D-3d}
{\rm D}\widetilde{\Psi}=  \frac 1 {(\rho^\#)^{\frac 32}}\left({\rm D}\Psi- \frac 32 \frac {{\rm d}{\cal M}} {{\cal M}} \Psi \right)\; , \qquad
\\ \label{DbPsi=4D-3d}
{\rm D}\bar{\widetilde{\Psi}}=  \frac 1 {(\rho^\#)^{\frac 32}}\left({\rm D}\bar{\Psi}- \frac 32 \frac {{\rm d}{\cal M}} {{\cal M}} \bar{\Psi} \right)\; , \qquad
\eea
where (see \eqref{om0=fqq+dM})
\bea\label{DZ=3d} {\rm D}{\bb Z}&=& {\rm d}{\bb Z}+ if^{qq}{\bb Z}+ [{\bb A}, {\bb Z}] \, , \qquad
\\ \label{DPsi=3d} {\rm D}\Psi&=& {\rm d}\Psi- \frac i 2 f^{qq}\Psi+ [{\bb A}, \Psi]\; . \qquad \eea

Now, writing the 4D nAmW action in terms of new matrix variables and reduced spinor frame variables
\eqref{rv-=v2-iv1}, and using the  new covariant derivatives \eqref{DZ=3d}, \eqref{DPsi=3d} and  supervielbein forms \eqref{E--=cME0+}--\eqref{E+=cME0-}, we arrive at
\begin{eqnarray}
\label{SmM0'=4D} S_{\rm nAmW}\vert_{\eqref{rv-=v2-iv1}} &=&
\int\limits_{{\cal W}^1} {\rm  E}^{0} \left({\cal M}+ \frac 1 {{\cal M}} \, \frac {{\cal H}}{\mu^6}   \right)+ \nonumber \\  && +
\int\limits_{{\cal W}^1}  \left(- {\cal M}+ \frac 1 {{\cal M}} \, \frac {{\cal H}}{\mu^6}   \right)\, ({\rm d}x^2 - {\rm d}\theta^\gamma \bar{\theta}_\gamma+  \theta^\gamma {\rm d}\bar{\theta}_\gamma ) + \nonumber \\
&&
+ {1\over \mu^6} \int\limits_{{\cal W}^1}
 {\rm tr}\left(\bar{\bb P}{\rm D} {\bb Z} + {\bb P}{\rm D} \bar{\bb Z} - {i\over 8} {\rm D}{ \Psi}\,  \bar{\Psi} + {i\over 8} { \Psi} {\rm D} \bar{\Psi}  \right) -  \quad \nonumber \\  && -
 {1\over \mu^6} \int\limits_{{\cal W}^1} \frac {{\rm d}{\cal M}} {{\cal M}}
 {\rm tr}\left(\bar{\bb P} {\bb Z} + {\bb P} \bar{\bb Z}\right)  +
\quad \nonumber \\  &&+ {1\over \mu^6} \int\limits_{{\cal W}^1}  \frac i {\sqrt{2}\sqrt{{\cal M}}}\left( {\cal E}^{1}-i{\cal E}^{2}\right)
 {\rm tr}(\bar{\Psi}  \bar{\bb P}+   \Psi [{\bb Z},  \bar{\bb Z}])+ \quad \nonumber \\  &&+ {1\over \mu^6} \int\limits_{{\cal W}^1}  \frac i {\sqrt{2}\sqrt{{\cal M}}}\left(\bar{ {\cal E}}^{1}+i\bar{{\cal E}}^{2}\right)
 {\rm tr}({\Psi}  {\bb P}+  \bar{\Psi} [{\bb Z},  \bar{\bb Z}]) . \qquad
\end{eqnarray}
Here ${\cal H}$ is of the same form as $\widetilde{{\cal H}}$ \eqref{HSYM=4D},  but written in terms of new $SO(1,1)$ invariant matrix variables,
\begin{eqnarray}
\label{tcH=} && {\cal H}=    {\rm tr}\left( {\bb P} \bar{\bb P} +  [{\bb Z},  \bar{\bb Z}]^2 -
{i\over 2} {\bb Z}{ \Psi}{ \Psi} + {i\over 2} \bar{\bb Z} \bar{\Psi}  \bar{\Psi} \right) \; ,
\qquad
\end{eqnarray}
and the real bosonic  ${\rm E}^0$ and complex fermionic ${\cal E}^q= (\bar{{\cal E}}^q)^*$ 1-forms are defined in \eqref{rmE0=} and \eqref{cEq=}, respectively.

As we have already discussed on simpler examples, the dimensional reduction procedure implies the use of the equations of motion for one of the bosonic coordinate functions, $x^2(\tau)$ in our case. This coordinate function enters only once, in the second term of \eqref{SmM0'=4D}, and its equation of motion
\be\label{x2=Eq}
{\rm d} \left({\cal M}- \frac 1 {{\cal M}} \, \frac {{\cal H}}{\mu^6}   \right)=0\;
\ee
implies
\be\label{cM-cH/cM=m}
{\cal M}- \frac 1 {{\cal M}} \, \frac {{\cal H}}{\mu^6} = m= \rm const\;
\ee
with constant $m$ of dimension of mass.
Eq. \eqref{cM-cH/cM=m} is solved by
\be\label{cM=m+-}
{\cal M}_\pm =\frac m 2\pm \sqrt{\frac {m^2} 4+\frac {{\cal H}}{\mu^6}}\; .
\ee
In our context only  the solution with plus sign makes sense, as it has nonvanishing value when
${{\cal H}}=0$ (in particular when all matrix field vanish),
\be\label{cM=m+}
{\cal M}= {\cal M} ({{\cal H}}/\mu^6) :=\frac m 2 + \sqrt{\frac {m^2} 4+\frac {{{\cal H}}}{\mu^6}}= m  \left[ 1 + \frac {{{\cal H}}}{m^2\mu^6} + {\cal O} \left( \left(\frac {{{\cal H}}}{m^2\mu^6}\right)^2\right) \right] \; .
\ee
The second equality gives the decomposition of the solution in power series in $\frac 1 {\mu^6}$. Notice that this power series can be also treated as weak field decomposition in the matrix fields, as decomposition of low energy of relative motion or as decomposition in $\frac 1 {m^2}$. Below we will refer on it  as on decomposition in coupling constant $\frac 1 {\mu^6}$ just for convenience.

Substituting the solution \eqref{cM=m+} of the $x^2$ equation of motion back to the action \eqref{SmM0'=4D} we find the following candidate action for the description of 3D mD$0$ system:
\begin{eqnarray}
\label{SmD0=3D} S_{\rm mD0}^{\rm 3D} &=& \int\limits_{{\cal W}^1} \left(m {\rm E}^{0}+ m ({\rm d}\theta^\gamma \bar{\theta}_\gamma- \theta^\gamma {\rm d}\bar{\theta}_\gamma )\right) + \nonumber \\
&
+& {1\over \mu^6} \int\limits_{{\cal W}^1}
 {\rm tr}\left(\bar{\bb P}{\rm D} {\bb Z} + {\bb P}{\rm D} \bar{\bb Z} - {i\over 8} {\rm D}{ \Psi}\,  \bar{\Psi} + {i\over 8} { \Psi} {\rm D} \bar{\Psi}  \right) -  {1\over \mu^6} \int\limits_{{\cal W}^1}
 \frac {{\rm d}{\cal M}} {{\cal M}}
 {\rm tr}\left(\bar{\bb P} {\bb Z}+ {\bb P} \bar{\bb Z}\right) +  \nonumber \\  &&   + {1\over \mu^6} \int\limits_{{\cal W}^1}  {\rm E}^{0}\frac 2 {{\cal M}} \, {{\cal H}}
   + {1\over \mu^6} \int\limits_{{\cal W}^1}   \frac i {\sqrt{2}\sqrt{{\cal M}}}\left( {\cal E}^{1}-i{\cal E}^{2}\right)
 {\rm tr}(\bar{\Psi}  \bar{\bb P}+   \Psi [{\bb Z},  \bar{\bb Z}])+ \quad \nonumber \\  &&+ {1\over \mu^6} \int\limits_{{\cal W}^1}  \frac i {\sqrt{2}\sqrt{{\cal M}}}\left(\bar{ {\cal E}}^{1}+i\bar{{\cal E}}^{2}\right)
 {\rm tr}({\Psi}  {\bb P}+  \bar{\Psi} [{\bb Z},  \bar{\bb Z}])  \qquad \;
\end{eqnarray}
with  ${\cal H}$ defined in \eqref{tcH=}, bosonic and fermionic 1-forms  defined in
\eqref{rmE0=} and \eqref{cEq=}, covariant derivatives defined in \eqref{DZ=3d}, \eqref{DPsi=3d}
and
${\cal M}= {\cal M} ({\cal H}/\mu^6)$ given in \eqref{cM=m+}.

Actually, as we will discuss below, the action \eqref{SmD0=3D} with arbitrary (nonvanishing) function ${\cal M} ({{\cal H}}/\mu^6)$ also makes sense.

Notice that, as the  derivative of
${\cal M}= {\cal M} ({\cal H}/\mu^6)$ (independently of whether it is given by  \eqref{cM=m+} or considered to be arbitrary)  is proportional to $\frac 1 {\mu^6}$ (actually to $\frac 1 {m^2\mu^6}$),
\be
{\rm d}{\cal M}= \frac 1 {\mu^6} \, \frac { {\rm d}{{\cal H}}} {2\sqrt{\frac {m^2} 4+\frac {{{\cal H}}}{\mu^6}}}\; ,
\ee
so that at the first order in $ \frac 1 {\mu^6} $ the action does not contain d${{\cal H}}$ and reads
\begin{eqnarray}
\label{SmD0=3D=} S_{\rm mD0}^{\rm 3D}\vert_{{\cal M}\mapsto m} &=& \int\limits_{{\cal W}^1} \left(m {\rm E}^{0}+ m ({\rm d}\theta^\gamma \bar{\theta}_\gamma- \theta^\gamma {\rm d}\bar{\theta}_\gamma )\right) + \nonumber \\
&
+& {1\over \mu^6} \int\limits_{{\cal W}^1}
 {\rm tr}\left(\bar{\bb P}{\rm D} {\bb Z} + {\bb P}{\rm D} \bar{\bb Z} - {i\over 8} {\rm D}{ \Psi}\,  \bar{\Psi} + {i\over 8} { \Psi} {\rm D} \bar{\Psi}  \right)  +  \nonumber \\  &&   + {1\over \mu^6} \int\limits_{{\cal W}^1}  {\rm  E}^{0}\frac 2 {m} \, {{\cal H}}
   + {1\over \mu^6} \int\limits_{{\cal W}^1}   \frac i {\sqrt{2}\sqrt{m}}\left( {\cal E}^{1}-i{\cal E}^{2}\right)
 {\rm tr}(\bar{\Psi}  \bar{\bb P}+   \Psi [{\bb Z},  \bar{\bb Z}])+ \quad \nonumber \\  &&+ {1\over \mu^6} \int\limits_{{\cal W}^1}  \frac i {\sqrt{2}\sqrt{m}}\left(\bar{ {\cal E}}^{1}+i\bar{{\cal E}}^{2}\right)
 {\rm tr}({\Psi}  {\bb P}+  \bar{\Psi} [{\bb Z},  \bar{\bb Z}]) \; . \qquad
\end{eqnarray}

\section{Worldline supersymmetry of the $D=3$ ${\cal N}=2$ mD$0$-brane action }

In this section we will show that the properties expected from mD$0$ system are possessed by a more  generic system described by the functional \eqref{SmD0=3D} with an arbitrary but nonvanishing function ${\cal M}= {\cal M} ({\cal H}/\mu^6)$. Namely we will show that such action possesses, besides the manifest $D=3$ ${\cal N}=2$ supersymmetry, also the worldline supersymmetry generalizing the $\kappa-$symmetry \eqref{kappa=D0} of single D$0$-brane action \eqref{SD0D3=}.

\subsection{Worldline supersymmetry transformations of the center of energy variables }

The previous experience  with 4D nAmW system and with 10D action of \cite{Bandos:2018ntt} suggests, when searching for worldline supersymmetry of mD$0$-brane,  to assume that it acts on the center of energy variables of the mD$0$ system, i.e. on the coordinate functions and spinor frame variables,  in the same manner as the $\kappa-$symmetry of single D$0$ brane,
\bea\label{kappa=mD0}
\delta_\epsilon \theta^\alpha =  \frac 1 {\sqrt{2}} ({\rm v}^{\alpha 2}- i{\rm v}^{\alpha 1})\, \epsilon\; , \qquad \delta_\epsilon \bar{\theta}{}^\alpha = \frac 1 {\sqrt{2}}  ({\rm v}^{\alpha 2}+ i{\rm v}^{\alpha 1})\, \bar{\epsilon}\; , \qquad \nonumber \\
\delta_\epsilon x^{\alpha\beta}= 2i \delta_\epsilon \theta^{(\alpha}  \bar{\theta}{}^{\beta)}-  2i  \theta^{(\alpha} \delta_\epsilon \bar{\theta}{}^{\beta)}\; , \qquad \nonumber \\
\delta_\epsilon {\rm  v}^q_\alpha =0\; .  \qquad
\eea

In the formalism of Appendix A this is
tantamount to stating that
\bea\label{kappa==}
i_\epsilon ({\cal E}^{1} +i {\cal E}^{2})=0 \; , \qquad i_\epsilon  (\bar{{\cal E}}{}^{1} -i \bar{{\cal E}}{}^{2})=0 \; , \qquad i_\epsilon {\rm E}^{0} =0\; , \qquad  i_\epsilon {\rm E}^{I} =0\; , \qquad \nonumber \\
i_\epsilon  f^I=0\; , \qquad i_\epsilon  f^{qq}=0\qquad \Rightarrow \qquad \delta_\epsilon {\rm v}_\alpha^\mp =0 \; , \qquad
\eea
and, consequently,
\bea\label{kappa=}
i_\epsilon  {\cal E}^{1} = \epsilon/\sqrt{2}  \qquad &\Rightarrow & \qquad i_{\epsilon}{\cal E}^{2} =\;  i\epsilon/\sqrt{2}  \; , \;\qquad  i_\epsilon ({\cal E}^1-i {\cal E}^2)= \sqrt{2} \epsilon  \; , \qquad \nonumber \\
i_\epsilon  \bar{{\cal E}}{}^{1} =\bar{\epsilon } /\sqrt{2} \qquad & \Rightarrow & \qquad   i_\epsilon  \bar{{\cal E}}{}^{2} =-i\bar{\epsilon }/\sqrt{2} \; , \qquad  i_\epsilon (\bar{{\cal E}}{}^1+i \bar{{\cal E}}{}^2)= \sqrt{2} \bar{\epsilon }  \; .  \qquad
\eea

In this form it is easier to find  the following transformation properties of the bosonic and fermionic 1-forms entering the part of the action containing the matrix fields
\bea
\label{1dSG=vE0}
\delta_{\epsilon }{\rm E}^0= -2i\frac {1}{\sqrt{2}} ({\cal E}^1-i {\cal E}^2)\bar{\epsilon } -2i\frac {1}{\sqrt{2}} (\bar{\cal E}{}^1+i \bar{\cal E}{}^2)\epsilon \; , \qquad \nonumber \\
\delta_\epsilon ({\cal E}^1-i {\cal E}^2)= \sqrt{2} {\rm D} \epsilon  \; , \qquad
\delta_\epsilon  (\bar{\cal E}{}^1+i \bar{\cal E}{}^2)=\sqrt{2} {\rm D}\bar{\epsilon } \; , \qquad
  \eea
which are the typical transformations of the $d=1$ ${\cal N}=2$ supergravity multiplet. This supergravity is induced by embedding of the worldline into $D=3$ ${\cal N}=2$ superspace  ({\it cf.} \eqref{wsSG4D} and discussion around it).

Eqs. \eqref{1dSG=vE0} are useful to search for the worldline supersymmetry invariance of the part of the action containing the matrix fields.

\subsection{Worldline supersymmetry transformations of the matrix matter fields }

Now, writing the variation of the action \eqref{SmD0=3D} with \eqref{kappa=mD0}--\eqref{1dSG=vE0} and
extracting from this variation the terms  proportional to D${\bb P}$,  D${\bb Z}$, D${\Psi}$, and their hermitian conjugates, we find the following {\it equations} for the basic $\kappa-$symmetry  variations of matrix `matter' fields
\bea\label{vkZ=}
\delta_\epsilon {\bb Z} = -\frac i {\sqrt{\cal M}}\epsilon \bar{\Psi} + \frac 1 {\mu^6} \, \frac{{\cal M}^\prime }{{\cal M}}\; \left({\bb Z} \delta_\epsilon {\tilde{{\cal H}}} - {\bb P} \Delta_\epsilon {\cal K} \right)\; ,
\\
\delta_\epsilon \bar{{\bb Z}} = -\frac i {\sqrt{\cal M}}\bar{\epsilon} \Psi + \frac 1 {\mu^6} \, \frac{{\cal M}^\prime }{{\cal M}}\; \left(\bar{{\bb Z}}  \delta_\epsilon {\tilde{{\cal H}}} -\bar{{\bb P}}  \Delta_\epsilon {\cal K} \right) \;,
\\
\delta_\epsilon {\bb P} =+\frac i {\sqrt{\cal M}}\left(\epsilon [\Psi,{{\bb Z}} ]+  \bar{\epsilon}  [\bar{\Psi},{\bb Z} ] \right) - \frac 1 {\mu^6} \, \frac{{\cal M}^\prime }{{\cal M}}\; {\bb P}  \delta_\epsilon {\tilde{{\cal H}}}~+ \qquad \nonumber  \\
 + \frac 1 {\mu^6} \, \frac{2{\cal M}^\prime }{{\cal M}}\; \left([[{\bb Z},\bar{{\bb Z}}],{\bb Z}]+\frac i 4 \bar{\Psi}\bar{\Psi}\right) \,  \Delta_\epsilon {\cal K}\;,  \\
\delta_\epsilon \bar{{\bb P}} = +\frac i {\sqrt{\cal M}}\left(\epsilon [\bar{{\bb Z}},\Psi ]+  \bar{\epsilon}  [\bar{{\bb Z}},\bar{\Psi} ] \right) - \frac 1 {\mu^6} \, \frac{{\cal M}^\prime }{{\cal M}}\; \bar{{\bb P}}  \delta_\epsilon {\tilde{{\cal H}}}~ - \qquad \nonumber \\
 - \frac 1 {\mu^6} \, \frac{2{\cal M}^\prime }{{\cal M}}\; \left([[{\bb Z},\bar{{\bb Z}}],\bar{{\bb Z}}]+\frac i 4 \Psi{\Psi}\right) \,  \Delta_\epsilon {\cal K}\;, \\
 \delta_\epsilon {\Psi} = +\frac 4 {\sqrt{\cal M}}\left(\epsilon  \bar{{\bb P}} + \bar{\epsilon} [{\bb Z}, \bar{{\bb Z}}] ] \right) - \frac 1 {\mu^6} \, \frac{2{\cal M}^\prime }{{\cal M}}\; [\bar{\Psi},\bar{{\bb Z}}]   \Delta_\epsilon {\cal K}\; , \\ \label{vkbPsi=}
 \delta_\epsilon \bar{{\Psi}} = +\frac 4 {\sqrt{\cal M}}\left(\epsilon [{\bb Z}, \bar{{\bb Z}}]+  \bar{\epsilon}  {\bb P}] \right) + \frac 1 {\mu^6} \, \frac{2{\cal M}^\prime }{{\cal M}}\; [\Psi,{\bb Z} ]   \Delta_\epsilon {\cal K}\; .
\eea
These are equations because their right hand sides contain variations
$\delta_{\epsilon} {{{\cal H}}}$ of $ {{{\cal H}}} $ from  \eqref{tcH=} and
\be
\Delta_\epsilon {\cal K}=\delta_\epsilon {\cal K} -\frac i{2\sqrt{{\cal M}}}\, (\epsilon\, {\nu}  +\bar{\epsilon}\bar{\nu})
\ee
where
\be
\nu = {\rm tr}\left(\bar{\Psi}\bar{{\bb P}} +\Psi[{\bb Z},\bar{{\bb Z}}]  \right)\; , \qquad  \bar{\nu}= {\rm tr}\left(\Psi{\bb P} +\bar{\Psi}[{\bb Z},\bar{{\bb Z}}]  \right)
\ee
and $\delta_{\epsilon} {\cal K}$ is the variation
 of
\be
{\cal K} ={\rm tr}(\bar{{\bb P}}{\bb Z}+{\bb P}\bar{{\bb Z}})\, .
\ee
As both these composite variations enters  with coefficients $\propto \frac 1 {\mu^6}$,
the equations certainly have the solution, at least  as power series in $\frac 1 {\mu^6}$.
But moreover,  there exists a simple way to solve these equations  which consists in firstly, using them to obtain closed algebraic equations for $\delta_{\epsilon} {{{\cal H}}}$ and $\delta_{\epsilon} {\cal K}$, secondly, solve these and, thirdly, substituting these solutions for the last terms in \eqref{vkZ=}--\eqref{vkbPsi=}.

Indeed,  using \eqref{vkZ=}--\eqref{vkbPsi=} we find
\be\label{vtcH=Eq}
 \delta_\epsilon  {{{\cal H}}} = \frac i{\sqrt{{\cal M}}}\, {\rm tr}\left[\left(\bar{\epsilon}\bar{\Psi}-\epsilon\Psi \right)\, \left([{\bb Z},\bar{{\bb P}}]+ [\bar{{\bb Z}},{\bb P}] -\frac i 4 \{\Psi\, , \bar{\Psi}\} \right)\right]   + \frac 1 {\mu^6} \, \frac {{\cal M}^\prime }{{\cal M}}\, {\frak{H} }\, \delta_\epsilon  {{{\cal H}}}
\ee
and
\be\label{vcK=Eq}
 \delta_\epsilon  {\cal K} = -\frac i{\sqrt{{\cal M}}}\, \epsilon\, {\rm tr}\left(\bar{\Psi}\bar{{\bb P}} -2\Psi[{\bb Z},\bar{{\bb Z}}]  \right)\, -\frac i{\sqrt{{\cal M}}}\, \bar{\epsilon}\, {\rm tr}\left(\Psi{\bb P} -2\bar{\Psi}[{\bb Z},\bar{{\bb Z}}]  \right)\,  + \frac 1 {\mu^6} \, \frac {{\cal M}^\prime }{{\cal M}}\, {\frak{H} }\, \Delta_\epsilon  {\cal K}\; ,
\ee
where
\be
\label{h=} {\frak{H}}= {\rm tr}\left(-2 {\bb P} \bar{\bb P} +  4 [{\bb Z},  \bar{\bb Z}]^2 -
{i\over 2} {\bb Z}{ \Psi}{ \Psi} + {i\over 2} \bar{\bb Z} \bar{\Psi}  \bar{\Psi} \right) \;
\qquad
\ee
 ({\it cf.} \eqref{tcH=}). Eqs. \eqref{vtcH=Eq} and \eqref{vcK=Eq} are closed algebraic equations for $\delta_\epsilon  {{{\cal H}}}$ and $\delta_\epsilon  {\cal K} $, respectively, which are solved by
\bea\label{vtcH=}
 \delta_\epsilon  {{{\cal H}}} &=& \frac i{\sqrt{{\cal M}}}\,\frac 1 {\left(1- \frac 1 {\mu^6} \, \frac {{\cal M}^\prime }{{\cal M}}\, {\frak{H} }\right)}\;   {\rm tr}\left[\left(\bar{\epsilon}\bar{\Psi}-\epsilon\Psi \right)\, \left([{\bb Z},\bar{{\bb P}}]+ [\bar{{\bb Z}},{\bb P}] -\frac i 4 \{\Psi\, , \bar{\Psi}\} \right)\right]
\eea
and
\bea\label{vcK=}
 \Delta_\epsilon  {\cal K} = -\frac {3i}{\sqrt{{\cal M}}}\, \frac 1 {\left(1- \frac 1 {\mu^6} \, \frac {{\cal M}^\prime }{{\cal M}}\, {\frak{H} }\right)}\; \left[ \epsilon\, {\rm tr}\left(\bar{\Psi}\bar{{\bb P}} -\Psi[{\bb Z},\bar{{\bb Z}}]  \right)\,+ \bar{\epsilon}\, {\rm tr}\left(\Psi{\bb P} -\bar{\Psi}[{\bb Z},\bar{{\bb Z}}]  \right)\right] \; .
\eea

Thus the worldline supersymmetry transformations of the matrix matter fields are given by  \eqref{vkZ=}$-$\eqref{vkbPsi=}  with  \eqref{vtcH=} and
\eqref{vcK=}.

\subsection{Worldline supersymmetry transformations of the non-Abelian gauge field }

Taking into account the above relations, we find that the remaining expression of the action variation contains the terms  proportional to the pull-backs of bosonic and fermionic supervielbein forms to the worldline, namely to
\be\label{E0cE0=1dSG}
{\rm E}^0\; ,\qquad ({\cal E}^1-i{\cal E}^2) \; ,\qquad  (\bar{{\cal E}}^1+i\bar{{\cal E}}^2)\;
\ee
and to the variation of $SU(N)$ gauge field 1-form $\delta {\bb A}$. This implies that, if the action is invariant under worldline supersymmetry, then
\be\label{vbbA=}
\delta_\epsilon {\bb A} =
{\rm E}^0\delta_\epsilon {\bb A} _0+({\cal E}^1-i{\cal E}^2)\delta_\epsilon {\bb A}_\eta + (\bar{{\cal E}}^1+i\bar{{\cal E}}^2)\delta_\epsilon {\bb A}_{\bar{\eta}}
\; .\qquad \ee

It is important to notice that $\delta_\epsilon {\bb A}$ enters the action variation in the trace of its product with $\left([{\bb Z},\bar{{\bb P}}]+ [\bar{{\bb Z}},{\bb P}] -\frac i 4 \{\Psi\bar{\Psi}\} \right)$,
\be
 {\rm tr}\left[\delta_\epsilon {\bb A}\, \left([{\bb Z},\bar{{\bb P}}]+ [\bar{{\bb Z}},{\bb P}] -\frac i 4 \{\Psi,\bar{\Psi}\} \right)\right]\; ,
\ee
so that the possibility to compensate all remaining terms in the action variation by choosing appropriate
$\delta_\epsilon {\bb A}_0$, $\delta_\epsilon {\bb A} _\eta$ and $\delta_\epsilon {\bb A} _{\bar{\eta}}$ is a nontrivial check of consistency of our calculations.

For instance, this implies that the condition of vanishing the contribution  $\propto {\rm E}^0$ in the $\kappa-$symmetry variation of the action,
\be
{\rm tr}\left[\delta_\epsilon {\bb A}_0\, \left([{\bb Z},\bar{{\bb P}}]+ [\bar{{\bb Z}},{\bb P}] -\frac i 4 \{\Psi,\bar{\Psi}\} \right)\right] = -\frac 2 {{\cal M}}\, \left(1-\frac 1 {\mu^6} \, \frac {{\cal M}^\prime }{{\cal M}}\right)\, \delta_\epsilon  {{\cal H}}\; ,
\ee
can be solved because  $\delta_\epsilon {\cal H}$ \eqref{vtcH=} is also given by the trace of certain expression with $\left([{\bb Z},\bar{{\bb P}}]+ [\bar{{\bb Z}},{\bb P}] -\frac i 4 \{\Psi ,\bar{\Psi}\} \right)$. This allows to obtain
\bea
\delta_\epsilon {\bb A}_0 &=&
 - \frac  {\left(1- \frac 1 {\mu^6} \, \frac {{\cal M}^\prime }{{\cal M}}\, {\cal H}\right)} {\left(1- \frac 1 {\mu^6} \, \frac {{\cal M}^\prime }{{\cal M}}\, {\frak{H} }\right)}\;  \frac {2i}{{\cal M}\sqrt{{\cal M}}}\, \left(\bar{\epsilon}\bar{\Psi}-\epsilon\Psi \right)\; .  \qquad
 \eea

Similarly studying the terms $\propto ({\cal E}^1-i{\cal E}^2)$ and their c.c.-s   we obtain
\bea\label{Aeta=}
\delta_\epsilon {\bb A}_\eta &=&  \frac {8i}{\sqrt{2}{\cal M}}\, \epsilon \bar{{\bb Z}}- \frac 1 {\mu^6}\, \,  \frac {{\cal M}^\prime }{2\sqrt{2}{\cal M}^2}
  \left(2i\Psi \sqrt{{\cal M}}\Delta_\epsilon {\cal K}- \frac {  3\left(\epsilon\Psi -\bar{\epsilon}\bar{\Psi}\right) {\rm tr}\left(\bar{{\bb P}}\bar{\Psi} - \Psi [{\bb Z},\bar{{\bb Z}}]\right)}  {\left(1- \frac 1 {\mu^6} \, \frac {{\cal M}^\prime }{{\cal M}}\, {\frak{H} }\right)} \right) \,  \nonumber \\ &=&
   \frac {8i}{\sqrt{2}{\cal M}}\, \epsilon \bar{{\bb Z}}-   \frac 1 {\mu^6}\, \,  \frac {3{\cal M}^\prime }{2\sqrt{2}{\cal M}^2}\frac  {1} {\left(1- \frac 1 {\mu^6} \, \frac {{\cal M}^\prime }{{\cal M}}\, {\frak{H} }\right)} \times \qquad  \,  \nonumber \\
 && \hspace{2cm} \times
  \;  \left[ (\bar{\epsilon}\bar{\Psi}-2{\epsilon}{\Psi}) {\rm tr}\left(\bar{{\bb P}}\bar{\Psi} -  \Psi [{\bb Z},\bar{{\bb Z}}]\right)  +  \bar{\epsilon}\Psi {\rm tr}\left(\bar{{\bb P}}\bar{\Psi} - \,\Psi [{\bb Z},\bar{{\bb Z}}]\right)\right]\; \qquad
\eea
and
\bea
\delta_\epsilon {\bb A}_{\bar{\eta}} &=& - \frac {8i}{\sqrt{2}{\cal M}}\, \bar{\epsilon} {\bb Z}+ \frac 1 {\mu^6}\, \,  \frac {3{\cal M}^\prime }{2\sqrt{2}{\cal M}^2}
  \frac  {\epsilon\Psi -\bar{\epsilon}\bar{\Psi}} {\left(1- \frac 1 {\mu^6} \, \frac {{\cal M}^\prime }{{\cal M}}\, {\frak{H} }\right)}\;   {\rm tr}\left({\bb P}\Psi - \bar{\Psi} [{\bb Z},\bar{{\bb Z}}]\right) + \qquad \nonumber \\ &&
  + \frac 1 {\mu^6} \, \frac {i{\cal M}^\prime }{\sqrt{2}\sqrt{{\cal M}}{\cal M}}\;\bar{\Psi}\; \Delta_\epsilon {\cal K}  \, . \qquad
\eea
The second of these equations can be further specified with the use of   \eqref{vcK=}.

\subsection{Resume on the  candidate mD$0$ action(s)}

Thus  we have shown that the actions \eqref{SmD0=3D} with arbitrary nonvanishing function ${\cal M}({\cal H}/\mu^6)$ of the matrix field Hamiltonian \eqref{tcH=} possesses, besides the target superspace $D=3$ ${\cal N}=2$  supersymmetry, also local  worldline supersymmetry which generalizes the $\kappa-$symmetry of the single-D$0$-brane action. Hence any of these can be considered as a candidate for the role of $D=3$ counterpart of the multiple D$0$-brane (mD$0$) action.
A special representative of this family is the action with ${\cal M}({\cal H}/\mu^6)$ given in \eqref{cM=m+}, as this is obtained by dimensional reduction from  the $D=4$ counterpart of the mM$0$-brane system (4D nAmW action).
Generically the actions  \eqref{SmD0=3D} are essentially nonlinear but the simplest representation of the family with ${\cal M}({\cal H}/\mu^6)=m=~$const. This simplest case provide us with the counterpart of the 10D action considered as a candidate on the role of mD$0$ action in \cite{Bandos:2018ntt}. Our study suggests to search for more generic essentially nonlinear candidates on the role of 10D mD$0$ action, and this problem is presently under investigation.

\newpage

\section{Conclusion}

The main result of this paper is doubly supersymmetric (i.e. possessing both rigid target superspace supersymmetry and worldline supersymmetry) candidate action(s)  for the description of 3D counterpart of 10D multiple-D$0$-brane (mD$0$), \eqref{SmD0=3D}.
It includes all bosonic and fermionic  fields which are expected to be present in the 3D mD$0$ system, which can be restored from the very low energy gauge fixed description by the $U$(N) SYM action \cite{Witten:1995im} and the known actions for single super-D$p$-branes \cite{Cederwall:1996pv}$-$\cite{Bandos:1997rq}. Furthermore it is invariant under both the target superspace $D=3$ ${\cal N}=2$  rigid supersymmetry and local worldline supersymmetry generalizing the $\kappa-$symmetry of the  single super-D$0$-brane (in its irreducible form characteristic for the spinor moving frame formulation \cite{Bandos:2000tg}). This latter is important because it guarantees the preservation of a part of target space supersymmetry by the ground state of the system, and this is what is expected from the multiple D$0$-brane.

Curiously enough the action includes an arbitrary nonvanishing function ${\cal M}({\cal H})$ of the `relative motion Hamiltonian' ${\cal H}$ which in its turn is constructed from the bosonic and fermionic matrix fields of mD$0$ system. Generically, the action is essentially nonlinear. An exception is the case of
${\cal M}({\cal H})=m={\rm const}$ \eqref{SmD0=3D=} which is given by the sum of single D$0$-brane action (the variables of which is now describing the center of mass motion of the system) and the action of $d=1$ ${\cal N}=2$ SYM the supersymmetry of which is made local by coupling to the supergravity induced by the center of energy motion (see \eqref{1dSG=vE0}). The $D=10$ counterpart of such an action was considered in \cite{Bandos:2018ntt} where, in particular, its comparison with the multiple 0-brane action from \cite{Panda:2003dj} was carried out.
Our present study suggests the existence of other candidates on the role of 10D mD$0$ actions and we will turn to the problem of their construction and studying in a forthcoming publication.

One might wonder whether it is possible to write also the generalization of 4D nAmW action and of its 11D mM$0$ prototype with arbitrary function $\tilde{{\cal M}}(\tilde{{\cal H}})$. At least presently we do not know such actions possessing, besides target space supersymmetry, also the worldline supersymmetry. In the case of 3D  mD$0$-brane the possibility to write the doubly supersymmetric action with arbitrary function ${\cal M}({\cal H})$ was found in  a way occasionally: we first obtained the action with definite function \eqref{cM=m+} by dimensional reduction of the $D=4$ nAmW  (4D mM$0$)  action, and found that it is more convenient to check  for the presence of
worldline supersymmetry first without specification of the form of ${\cal M}({\cal H})$. Then we have found that the deserved worldline supersymmetry is actually present in the case of arbitrary invertible  ${\cal M}({\cal H})$.

The problem of choice of the function ${\cal M}({\cal H})$ which  leads to the true mD$0$-brane action should be addressed in a more general perspective of String/M-theory. It is tempting to use the T-duality (which was the main argument  for construction of bosonic actions in \cite{Myers:1999ps}) to make this choice. However, to this end we need also to have a complete, doubly supersymmetric action for multiple super-D$1$-brane (mD$1$) system. For the moment, in our $D=3$ case a special role is played by the action with ${\cal M}({\cal H})$ given in \eqref{cM=m+} because it is obtained from the $D=4$ action for a nAmW system \eqref{SmM0=4D}, the 4D counterpart of the  multiple M0-brane system \cite{Bandos:2012jz} which was also obtained for the first time in the present paper.

Let us also notice that, if we froze all the center of energy degrees of freedom in our mD$0$ action with an arbitrary invertible ${\cal M}({\cal H})$,  we find a family of nonlinear generalizations of the (first order) action for 1d SYM  which for our knowledge are new. In higher dimensions the known nonlinear generalizations of the non-Abelian gauge field action are quite a few,  to the best of our knowledge, all related to the symmetric trace  BI action by Tseytlin \cite{Tseytlin:1997csa}. The study of the properties of  new 1d nonlinear supersymmetric gauge field models and their  applications looks interesting on its own.

The other interesting directions for the development of the line of present study is, first of all, the  already mentioned search for 10D generalization of the action \eqref{SmD0=3D}, proving its worldline supersymmetry, studying its properties  and its possible relation with 11D mM0-action of \cite{Bandos:2012jz}, as well as  addressing the problem of choice of the best candidate for the role of 10D mD$0$-brane action. This latter issue is related to the problem of the search for higher $p$ multiple D$p$-brane actions, beginning from $p=1$.

As far as the study of $D=3$ and $D=4$ supersymmetric systems is concerned, an interesting problem is the development of Hamiltonian approach and quantization of  $D=4$ ${\cal N}=1$ nAmW system as well as of $D=3$ ${\cal N}=2$ mD$0$-brane. These should result in new interesting supersymmetric systems of relativistic wave equations in the (super)spaces with noncommutative  coordinates which  promise to be simpler than the result of quantization of    $D=3$ ${\cal N}=1$ nAmW system in \cite{Bandos:2018qqo} due to the complex structure characteristic for the both $D=4$ ${\cal N}=1$  and $D=3$ ${\cal N}=2$ superspaces.

As far as quantization of  $D=4$ nAmW system is concerned, it can be considered as a preliminary step to approach the quantum description  of 11D mM0 system which in its turn might shed a new light on the properties of String/M-theory. In $D=3$ ${\cal N}=2$ case especially intriguing  looks the question on the influence of the choice of the function ${\cal M}({\cal H})$ on the results of the quantization.

\subsection*{Acknowledgements}
 Work of IB has also been partially supported by the Spanish MICINN/FEDER (ERDF EU) grant PGC2018-095205-B-I00,  by the Basque Government Grant IT-979-16, and by the Basque Country University program UFI 11/55.

\newpage

\appendix

\renewcommand{\theequation}{A.\arabic{equation}}
\setcounter{equation}0
\section{Notice on differential forms and variations }

\label{diffF}

In this Appendix we present some basic equations of the  differential form formalism.

Let $\Xi_q$ and $\Xi^\prime_p$ be differential $q$-form and $p$-form in a superspace with coordinates $Y^{\frak{M}}$. This is to say
\be\label{Xiq=}
\Xi_q= \frac 1 {q!} {\rm d}Y^{{\frak{M}}_q}\wedge \ldots \wedge {\rm d}Y^{{\frak{M}}_1}\Xi_{{\frak{M}}_1\ldots {\frak{M}}_q}(Y)\; , \qquad \Xi^\prime_p= \frac 1 {p!} {\rm d}Y^{{\frak{M}}_p}\wedge \ldots \wedge {\rm d}Y^{{\frak{M}}_1}\Xi^\prime_{{\frak{M}}_1\ldots {\frak{M}}_p}(Y)\; , \qquad
\ee
where $\wedge $ is the exterior product of the forms,
\be
{\rm d}Y^{{\frak{M}}}\wedge {\rm d}Y^{{\frak{N}}}= - (-1)^{\varepsilon ({\frak{M}})\,\varepsilon ({\frak{N}})} {\rm d}Y^{{\frak{N}}}\wedge {\rm d}Y^{{\frak{M}}}\; , \qquad Y^{{\frak{M}}}\, Y^{{\frak{N}}}=  (-1)^{\varepsilon ({\frak{M}})\,\varepsilon ({\frak{N}})} Y^{{\frak{N}}}\,Y^{{\frak{M}}}\;  \qquad
\ee
and $\varepsilon ({\frak{M}}) $ is the Grassmann parity of $Y^{{\frak{M}}}$.
For instance, in the case of $D=4$ ${\cal N}=1$ superspace, $Y^{{\frak{M}}}\mapsto Z^M= (x^\mu, \theta^\alpha, \bar{\theta}{}^{\dot{\alpha}})$ \eqref{ZM=} and
\be\label{vex=}
\varepsilon (x^\mu)=0, \qquad \varepsilon (\theta^\alpha )=1 , \qquad \varepsilon (\bar{\theta}{}^{\dot{\alpha}})=1\; . \qquad
\ee
For bosonic $p$- and $q$-forms,
\be
\Xi_q\wedge \Xi_p = (-1)^{pq} \Xi_p\wedge \Xi_q \; \qquad {\rm when}\qquad  \varepsilon (\Xi_q)=0 \; .
\ee
In the wedge product of two fermionic forms an additional minus sign appears, so that, e.g.
\be
{\rm d}\theta^\alpha \wedge {\rm d}\theta^\beta =  {\rm d}\theta^\beta  \wedge {\rm d}\theta^\alpha \; , \qquad  {\rm d}x^\mu \wedge {\rm d}\theta^\alpha = - {\rm d}\theta^\alpha  \wedge {\rm d}x^\mu \; , \qquad  {\rm d}x^\mu \wedge {\rm d}x^\nu = - {\rm d}x^\nu  \wedge {\rm d}x^\mu \; .
\ee

The exterior derivative of the differential forms is defined by
\bea\label{dXiq=}
{\rm d}\Xi_q&=& \frac 1 {q!} {\rm d}Y^{{\frak{M}}_q}\wedge \ldots \wedge {\rm d}Y^{{\frak{M}}_1} \wedge {\rm d}Y^{{\frak{M}}_0}
\partial_{{\frak{M}}_0}  \Xi_{{\frak{M}}_1\ldots {\frak{M}}_q}(Y) \qquad \nonumber \\ &\equiv &
\frac 1 {(q+1)!} {\rm d}Y^{{\frak{M}}_{q+1}}\wedge \ldots \wedge {\rm d}Y^{{\frak{M}}_2} \wedge {\rm d}Y^{{\frak{M}}_1}
(q+1) \partial_{[{\frak{M}}_2}  \Xi_{{\frak{M}}_2\ldots {\frak{M}}_{q+1}\}}(Y)\; . \qquad
\eea
Here the mixed brackets $[\ldots \}$ denote graded antisymmetrization over the enclosed indices, which implies symmetrization for exchanging the pair of fermionic indices and antisymmetrization for the pair including at least one bosonic index. The exterior derivative is nilpotent
\be
{\rm d}{\rm d}=0
\ee
and acts on the product of differential forms according to the following Leibniz rules
\be
{\rm d}(\Xi_p \wedge \Xi^\prime_q) =\Xi_p \wedge {\rm d}\Xi^\prime_q + (-)^q {\rm d}\Xi_p \wedge \Xi^\prime_q\; .
\ee

The variations of differential forms can be calculated with the use of Lie derivative formula,
\be
\delta \Xi_q = i_\delta {\rm d}\Xi_q +{\rm d} i_\delta\Xi_q\; , \qquad
\ee
where $i_\delta$ is the contraction with variation symbol defined by
\be
i_\delta \Xi_q= \frac 1 {(q-1)!} {\rm d}Y^{{\frak{M}}_q}\wedge \ldots \wedge {\rm d}Y^{{\frak{M}}_2} \delta Y^{{\frak{M}}_1} \Xi_{{\frak{M}}_1\ldots {\frak{M}}_q}(Y)\; , \qquad
\ee
 and hence obeying
\be
i_\delta(\Xi_p \wedge \Xi^\prime_q) =\Xi_p \wedge i_\delta\Xi^\prime_q + (-)^q i_\delta\Xi_p \wedge \Xi^\prime_q\; .
\ee

To use  the Lie derivative expression for  the variation of the Lagrangian D-form ${\cal L}$ in D-dimensional field theory,
\be\label{varL}
\delta {\cal L}= i_\delta ({\rm d}{\cal L}) +  {\rm d}(i_\delta {\cal L}) \; ,
\ee
the exterior derivative should be considered as formally taken in the space of more dimensions: better in the space where all the fields are treated on the equal footing with coordinates (coordinate-field democracy).

In particular  in the case of superparticle Lagrangian, to obtain its variation from the Lie derivative formula \eqref{varL}, the exterior derivative should be calculated  with considering all the 1d fields, $x^\mu (\tau)$, $v^-_\alpha (\tau) $ etc.,  to be replaced by independent variables $x^\mu$, $v^-_\alpha$ etc. rather than being the  functions of proper time $\tau$ only. Also the second term is the total derivative and hence is not essential when we derive the equations of motion.

Often it is convenient to use the covariant Lie derivative formula, e.g.
\be\label{varL=D}
\delta {\cal L}= i_\delta ({\rm D}{\cal L}) +  {\rm D}(i_\delta {\cal L}) \; ,
\ee
which is equivalent to \eqref{varL=D} in the case when connection in the covariant derivative corresponds to gauge transformations which leave invariant the Lagrangian form.

Examples of the use of simple and covariant Lie derivative formula are
\be\label{vPi}
{\rm d}\Pi^{\alpha\dot{\alpha}} = -4i {\rm d}\theta^\alpha \wedge d\bar{\theta}{}^{\dot\alpha} \qquad \Rightarrow \qquad
\delta \Pi^{\alpha\dot{\alpha}} = +4i \delta\theta^\alpha  {\rm d}\bar{\theta}{}^{\dot\alpha} -4i {\rm d}\theta^\alpha \delta \bar{\theta}{}^{\dot\alpha}+ {\rm d}(i_\delta \Pi^{\alpha\dot{\alpha}}) \; , \qquad \ee
as well as
\bea\label{vE++}
{\rm D}E^{\#}= -4i E^+\wedge\bar{E}{}^+  - E^{-+}\wedge \Omega^{++} - E^{+-}\wedge \bar{\Omega}^{++}  \qquad \Rightarrow \hspace{4cm} \nonumber \\ \Rightarrow  \quad \delta E^{\#}= -4i E^+ i_\delta \bar{E}{}^++ 4i i_\delta E^+ \bar{E}{}^+ + \left(
i_\delta E^{-+} \Omega^{++}- E^{-+}i_\delta \Omega^{++}+ \rm c.c.
\right) \; , \qquad \\ \label{vE+}
{\rm D}E^+= -E^-\wedge \Omega^{++} \qquad \Rightarrow \qquad \delta E^+= i_\delta E^-\Omega^{++} -E^-i_\delta \Omega^{++}+ {\rm D}i_\delta E^+\; , \qquad \\ \label{vbE+}
{\rm D}\bar{E}{}^+= -\bar{E}{}^-\wedge \bar{\Omega}{}^{++} \qquad \Rightarrow \qquad \delta \bar{E}{}^+= i_\delta \bar{E}^-\bar{\Omega}{}^{++} -\bar{E}{}^-i_\delta \bar{\Omega}{}^{++}+ {\rm D}i_\delta \bar{E}{}^+\; , \qquad  \eea and
\bea
\label{vE--}
{\rm D}E^{=}= -4i E^-\wedge\bar{E}{}^-  + E^{+-}\wedge \Omega^{--} + E^{-+}\wedge \bar{\Omega}^{--}  \qquad \Rightarrow \hspace{4cm} \nonumber \\ \Rightarrow  \quad \delta E^{=}= -4i E^- i_\delta \bar{E}{}^-+ 4i i_\delta E^- \bar{E}{}^- + \left(
-i_\delta E^{+-} \Omega^{--}+ E^{+-}i_\delta \Omega^{--}+ \rm c.c.
\right)
\; ,  \qquad \\
\label{vE-}
{\rm D}E^-= +E^+\wedge \Omega^{--} \qquad \Rightarrow \qquad \delta E^-=- i_\delta E^+\Omega^{--} +E^+i_\delta \Omega^{--}+ {\rm D}i_\delta E^-\; , \qquad \\ \label{vbE-}
{\rm D}\bar{E}{}^-= +\bar{E}{}^+\wedge \bar{\Omega}{}^{--} \qquad \Rightarrow \qquad \delta \bar{E}{}^-= -i_\delta \bar{E}^+\bar{\Omega}{}^{--}+\bar{E}{}^+i_\delta \bar{\Omega}{}^{--}+ {\rm D}i_\delta \bar{E}{}^-\; . \qquad
\eea

To illustrate the above formalism let us use it to obtain equations of motion for the  massless  superparticle  in $D=4$ ${\cal N}=1$ superspace. The formal exterior derivative of its Lagrangian form
\be
{\cal L}^{0\; \rm D=4} = \rho^\# E^= = \rho^\# \Pi^{\alpha\dot\beta}v_\alpha^-\bar{v}_{\dot{\beta}}^- \;
\ee
 is
\be
{\rm d}{\cal L}^{0\; \rm D=4} =- {\rm D} \rho^\#\wedge  E^= -4i\rho^\# E^-\wedge\bar{E}{}^-  + \rho^\#E^{+-}\wedge \Omega^{--} +\rho^\# E^{-+}\wedge \bar{\Omega}^{--} \; .
\ee
From this, using \eqref{varL}, we find
\bea
\delta {\cal L}^{0\; \rm D=4}& =&  i_\delta {\rm D}\rho^\# E^= - {\rm D} \rho^\# i_\delta E^=   -4i\rho^\# E^-i_\delta\bar{E}{}^-  +4i\rho^\# i_\delta E^-\bar{E}{}^-  + \nonumber \\  && + \rho^\# E^{+-} i_\delta\Omega^{--} - \rho^\#  i_\delta E^{+-} \Omega^{--} +\rho^\#  E^{-+} i_\delta\bar{\Omega}^{--} -\rho^\#  i_\delta E^{-+}\bar{\Omega}^{--}
\eea
which allows to easily find that the nontrivial equations of motion of the superparticle  in the spinor moving frame formulations:
\bea\label{E--=0}
& \frac {\delta {\cal L}^{0\; \rm D=4} }{ i_\delta {\rm D}\rho^\# }= \frac {\delta {\cal L}^{0\; \rm D=4} }{ \delta \rho^\# }=0\qquad & \Rightarrow \qquad  E^{=}=0\; , \qquad \\ \label{E+-=0}
& \frac {\delta {\cal L}^{0\;\rm D=4} }{ i_\delta \Omega^{--} }= v_\alpha^+ \frac {\delta {\cal L}^{0\;\rm D=4} }{ \delta v_\alpha^{-}}=0\qquad & \Rightarrow \qquad  \rho^\#  E^{+-}=0\; , \qquad \\
\label{E-+=0}
 & \frac {\delta {\cal L}^{0\;\rm D=4} }{ i_\delta \bar{\Omega}{}^{--} }= \bar{v}_{\dot{\alpha}}^+ \frac {\delta {\cal L}^{0\; \rm D=4} }{ \delta \bar{v}_{\dot{\alpha}}^{-}}=0\qquad & \Rightarrow \qquad  \rho^\#  E^{-+}=0\; , \qquad \\
 \label{Dr++=0}
 & \frac {\delta {\cal L}^{0\;\rm D=4} }{ i_\delta E^{=} }= \frac 1 2 u_{\mu}^\# \frac {\delta {\cal L}^{0\;\rm D=4} }{ \delta x^{\mu}}=0\qquad & \Rightarrow \qquad  {\rm D} \rho^\#  =0\; , \qquad \\
 \label{Om--=0}
 & \frac {\delta {\cal L}^{0\;\rm D=4} }{ i_\delta E^{+-} }= \frac 1 2 u_{\mu}^{-+} \frac {\delta {\cal L}^{0\;\rm D=4} }{ \delta x^{\mu}}=0\qquad & \Rightarrow \qquad  \rho^\#  \Omega^{--}=0\; , \qquad  \\  \label{bOm--=0}
 & \frac {\delta {\cal L}^{0\;\rm D=4} }{ i_\delta E^{-+} }= \frac 1 2 u_{\mu}^{+-} \frac {\delta {\cal L}^{0\;\rm D=4} }{ \delta x^{\mu}}=0\qquad & \Rightarrow \qquad  \rho^\#  \bar{\Omega}^{--}=0\; , \qquad  \\ \label{bE-=0}
 & \frac {\delta {\cal L}^{0\;\rm D=4} }{ i_\delta E^{-} }= -v^{\alpha +}  \frac {\delta {\cal L}^{0\;\rm D=4} }{ \delta \theta^{\alpha}}=0\qquad & \Rightarrow \qquad  \rho^\#  \bar{E}^{-}=0\; , \qquad  \\ \label{E-=0}
 & \frac {\delta {\cal L}^{0\;\rm D=4} }{ i_\delta\bar{E}^{-} }=-\bar{v}{}^{\dot{\alpha} +} \frac {\delta {\cal L}^{0\; \rm D=4} }{ \delta \bar{\theta}{}^{\dot{\alpha}}}=0\qquad & \Rightarrow \qquad  \rho^\#  {E}^{-}=0\; . \qquad
\eea

\section{$D=3$ massless superparticle by
dimensional reduction of spinor moving frame sector of the $D=4$ massless superparticle}
\renewcommand{\theequation}{B.\arabic{equation}}
\setcounter{equation}0

In this appendix we  discuss the derivation of the $D=3$ massless superparticle action from its $D=4$ counterpart in spinor frame formulation. This can be considered as an illuminating warm up exercise suggesting the way of  dimensional reductions of more complex systems discussed in the main text.

\bigskip

\subsection{Brink-Schwarz formulation}

\medskip

In the Brink-Schwarz formulation of the massless superparticle action \eqref{SBS=4D}
the dimensional reduction  can be performed by, firstly, separating, say $p_2$ component (we make this choice for future convenience),
\be
p_\mu = (p_{\tilde{\mu}}, p_2)\; , \qquad {\tilde{\mu}}=0,1,3 \qquad \leftrightarrow \qquad \mu= 0,1,2,3\; ,
\ee
and setting this component equal to zero,
\be\label{p2=0}
p_2=0\; .
\ee
Substituting \eqref{p2=0} as an ansatz into \eqref{SBS=4D}, we arrive at the action for $D=3$ massless superparticle,  \be\label{SBS=3D}
S_{\rm BS}^{\rm D=3}= \int \left( p_{\tilde{\mu}} \Pi^{\tilde{\mu}} +  \frac 1 2 {\rm d} \tau e p_{\tilde{\mu}} p^{\tilde{\mu}}\right)\; .
\ee

\bigskip

\subsection{Spinor moving frame  formulation}

\medskip

As in the spinor moving frame formulation the role of additional momentum variable are taken by bilinear of bosonic spinors,
$v_\alpha^-$ and $\bar{v}_{\dot{\alpha}}^-$, it is natural to expect that similar dimensional reduction can be performed just by choosing a suitable ansatz for the $D=4$  harmonics in terms of $D=3$ ones.

Actually the ansatz consists in imposing reality conditions
\be\label{v=v*}
v_\alpha^-=\bar{v}_{\dot{\alpha}}^- \; , \qquad v_\alpha^+=\bar{v}_{\dot{\alpha}}^+\; .
\ee
This choice is invariant under $SL(2,{\bb R})=Spin(1,2)$ subgroup of $SL(2,{\bb C})=Spin(1,3)$.
It is convenient to complete \eqref{v=v*} to
\be\label{v=v*=}
v_\alpha^-=\bar{v}_{\dot{\alpha}}^-={\rm v}_\alpha^- \; , \qquad v_\alpha^+=\bar{v}_{\dot{\alpha}}^+={\rm v}_\alpha^+\; ,
\ee
were we have introduced real  3D spinor moving frame variables ${\rm v}_\alpha^\pm$ which form the $SL(2,{\bb R})$ valued matrix, i.e. obey
\be\label{v-v+=1=3D*}
{\rm v}^{-\alpha}{\rm v}_\alpha^+=1 \; , \qquad ({\rm v}_\alpha^\pm)^*= {\rm v}_\alpha^\pm\; .
\ee

Using the split of the  $D=4$ relativistic Pauli matrices \eqref{Pauli=} as in
\eqref{s4d=s3d} and identifying the $D=3$ gamma matrices as in \eqref{g3=s4}, \eqref{tg3=ts4} we find that the ansatz \eqref{v=v*=} implies
\be\label{u2--=0}
u_2^==0\; , \qquad u_2^\#=0\; , \qquad u_2^{-+} = - u_2^{+-}=(u_2^{+-})^*\; ,
\ee
while the real three vectors formed from $0,1,3$ components of the 4D moving frame vectors
are identified with the components of the 3D moving frame vectors
\be
 {\rm u}_{\tilde{\mu}}^= = {\rm v}^-\gamma_{\tilde{\mu}} {\rm v}^- \equiv  {\rm v}^{-\alpha}\gamma_{\tilde{\mu}\, \alpha\beta} {\rm v}^{-\beta} \, , \qquad {\rm u}_{\tilde{\mu}}^\# =  {\rm v}^+\gamma_{\tilde{\mu}} {\rm v}^+ \, , \qquad
u_{\tilde{\mu}}^\perp =  {\rm v}^+\gamma_{\tilde{\mu}} {\rm v}^- \,   \qquad
\ee
as follows
\be
u_{\tilde{\mu}}^= = {\rm u}_{\tilde{\mu}}^=  \, , \qquad u_{\tilde{\mu}}^\# = {\rm u}_{\tilde{\mu}}^\#  \, , \qquad
u_{\tilde{\mu}}^{-+} = u_{\tilde{\mu}}^{+-}=u_{\tilde{\mu}}^\perp  \,  . \qquad
\ee

Now, let us turn to the 4D massless superparticle action \eqref{S0D4==}. With the ansatz
\eqref{v=v*=}, which results in \eqref{u2--=0}, one of four VA  1-forms,
\be \label{Pi2==}
\Pi^2= \text{d}x^2-i {\rm d}\theta\sigma^2\bar{\theta}+i\theta\sigma^2 {\rm d}\bar{\theta}= {\rm d}x^2-\epsilon_{\alpha\beta} ( {\rm d}\theta^{\alpha} \bar{\theta}^{\beta} -\theta^{\alpha} {\rm d}\bar{\theta}^{\beta})\; ,
\ee
just disappears from the action which becomes  3D massless superparticle action in its spinor moving frame formulation
\be\label{S0D3=}
S^0_{\rm D=3}= \int \rho^{\#} {\rm v}_\alpha^-\bar{{\rm v}}_{\beta}^- \Pi^{\alpha\beta}=
\int \rho^{\#} {\rm u}_{\tilde{\mu}}\Pi^{\tilde{\mu}}= \int \rho^{\#} {\rm E}^{=}\; .
\ee
Here we have introduced 3D Volkov-Akulov (VA)
 1-forms
\be\label{VA=3d=}
\Pi^{\alpha\beta}= \Pi^{\tilde{\mu}}\tilde{\gamma}_{\tilde{\mu}}^{\alpha\beta}= {\rm d}x^{\alpha\beta}- 2i {\rm d}\theta^{( \alpha}\bar{\theta}{}^{\beta )} + 2i \theta^{( \alpha}{\rm d}\bar{\theta}{}^{\beta )}\; , \qquad
\Pi^{\tilde{\mu}}={\rm d}x^{\tilde{\mu}}-i {\rm d}\theta{\gamma}^{\tilde{\mu}}\bar{\theta} + i \theta{\gamma}^{\tilde{\mu}}{\rm d}\bar{\theta}\; ,
\ee
as well as one of 3D counterparts of the pull-backs of the 4D supervielbein forms \eqref{E--=}--\eqref{E+-=} adapted to the embedding,
\bea
{\rm E}^{=}&=&   \Pi^{\tilde{\mu}}{\rm u}_{\tilde{\mu}}^= =  \frac 1 2  {\rm u}_{\alpha\beta}^= \Pi^{\alpha\beta} =
 \Pi^{\alpha\beta}{\rm v}_{\alpha}^-{\rm v}_{\beta}^-
\, , \qquad \\
{\rm E}^{\#}&=&   \Pi^{\tilde{\mu}}{\rm u}_{\tilde{\mu}}^\# =  \frac 1 2  {\rm u}_{\alpha\beta}^\# \Pi^{\alpha\beta} =
 \Pi^{\alpha\beta}{\rm v}_{\alpha}^+{\rm v}_{\beta}^+
\, , \qquad \\
{\rm E}^{\perp}&=&   \Pi^{\tilde{\mu}}{\rm u}_{\tilde{\mu}}^\perp =  \frac 1 2  {\rm u}_{\alpha\beta}^\perp \Pi^{\alpha\beta} =
 \Pi^{\alpha\beta}{\rm v}_{(\alpha}^-{\rm v}_{\beta)}^+
\, . \qquad
\eea

\section{Some properties and applications of $SL(2,{\bb R})/SO(2)$ Cartan forms }
\renewcommand{\theequation}{C.\arabic{equation}}
\setcounter{equation}0

Inverting the relations \eqref{u1du2} and \eqref{u0duI}  we find
\be
\left.\begin{matrix}{\rm d}{\rm u}^0_{\alpha\beta}= {\rm u}^1_{\alpha\beta}f^1 +{\rm u}^2_{\alpha\beta}f^2 \; ,  \cr
{\rm d}{\rm u}^1_{\alpha\beta}= {\rm u}^0_{\alpha\beta}f^1 +{\rm u}^2_{\alpha\beta}f^{qq}  , \cr  {\rm d}{\rm u}^2_{\alpha\beta}= {\rm u}^0_{\alpha\beta}f^2 -{\rm u}^1_{\alpha\beta}f^{qq} .  \cr
\end{matrix}\right\} \qquad  \Longleftrightarrow \qquad \begin{cases} \text{d}{\rm u}^0_{\alpha\beta}= {\rm u}^1_{\alpha\beta}f^1 +{\rm u}^2_{\alpha\beta}f^2 \; ,  \cr
{\rm D}{\rm u}^1_{\alpha\beta}={\rm d}{\rm u}^1_{\alpha\beta}-{\rm u}^2_{\alpha\beta}f^{qq}= {\rm u}^0_{\alpha\beta}f^1   , \cr  {\rm D}{\rm u}^2_{\alpha\beta}= {\rm d}{\rm u}^2_{\alpha\beta}+ {\rm u}^1_{\alpha\beta}f^{qq}= {\rm u}^0_{\alpha\beta}f^2 .  \cr
\end{cases}
\ee
and
\be
f^{11}=\frac 1 2 f^{qq}-\frac 1 2 f^{1}\; , \qquad f^{22}=\frac 1 2 f^{qq}+\frac 1 2 f^{1}\; , \qquad f^{12}=\frac 1 2 f^{2}\; .\qquad
\ee

Notice also that
\be
{\rm d}f^{qp}=-\epsilon_{q'p'}f^{qq'}\wedge f^{pp'}= f^{q}{}_{p'}\wedge f^{pp'}\; .
\ee
In particular this implies
\bea
{\rm d}f^{qq}=-\epsilon_{q'p'}f^{qq'}\wedge f^{qp'}= 2(f^{11}-f^{22})\wedge f^{12}\; , \qquad \\ {\rm d}(f^{11}-f^{22})=2f^{qq}\wedge f^{12}\; , \qquad \\ {\rm d}f^{12}= f^{11}\wedge f^{22} = -\frac 1 2 f^{qq}\wedge (f^{11}-f^{22})\; . \qquad
\eea
Equivalently, we can write this relations as
\be
{\rm D}f^1={\rm d}f^1 +f^{qq}\wedge f^2=0\; , \qquad {\rm D}f^2={\rm d}f^3 -f^{qq}\wedge f^1=0\; , \qquad {\rm d}f^{qq} = -f^{1}\wedge f^2\; . \qquad
\ee
In terms of covariant derivatives and $SO(2)$ Cartan forms the derivatives of 3d spinor harmonics are
\be\label{Dv1==}
{\rm D}{\rm v}_\alpha^1 ={\rm d}{\rm v}_\alpha^1+\frac 1 2 {\rm v}_\alpha^2f^{qq} = \frac 1 2 {\rm v}_\alpha^1f^{2} +
\frac 1 2 {\rm v}_\alpha^2f^{1}\; , \qquad {\rm D}{\rm v}_\alpha^2 ={\rm d}{\rm v}_\alpha^2-\frac 1 2 {\rm v}_\alpha^1f^{qq} = +
\frac 1 2 {\rm v}_\alpha^1f^{1} -\frac 1 2 {\rm v}_\alpha^2f^{2}  \; . \qquad
\ee

Using the above equations we can easily find
\be  {\rm D}{\cal E}^1 =\frac 1 2 {\cal E}^1\wedge f^{2}+ \frac 1 2 {\cal E}^2\wedge f^{1}\; , \qquad {\rm D}{\cal E}^2 =\frac 1 2 {\cal E}^1\wedge f^{1}- \frac 1 2 {\cal E}^2\wedge f^{2}\; , \qquad\ee
which imply
\bea
{\rm D}({\cal E}^1-i{\cal E}^2)  =\frac 1 2 ({\cal E}^1+i{\cal E}^2)\wedge (f^{2}-if^{1})\; , \qquad \\
{\rm D}(\bar{{\cal E}}^1+i\bar{{\cal E}}^2)  =\frac 1 2 (\bar{{\cal E}}^1-i\bar{{\cal E}}^2)\wedge (f^{2}+if^{1})\; . \qquad\eea
These equations and  the Ricci identities
\bea
{\rm D}{\rm D}{\bb Z}=+ i f^1\wedge f^2 {\bb Z} + [{\bb F},{\bb Z}]\; , \qquad \\
{\rm D}{\rm D}\bar{{\bb Z}}=- i f^1\wedge f^2\bar{{\bb Z}} + [{\bb F},\bar{{\bb Z}}]\; , \qquad \\
{\rm D}{\rm D}{\Psi}=- \frac i 2 f^1\wedge f^2{\Psi} + [{\bb F},{\Psi}]\; , \qquad  \\
{\rm D}{\rm D}\bar{{\Psi}}=+ \frac i 2 f^1\wedge f^2\bar{{\Psi}} + [{\bb F},\bar{{\Psi}}]\; , \qquad
\eea
are useful in search for $\kappa-$symmetry of the 3D mD$0$ action.

The equations
\be
{\rm d}{\rm E}^0 = {\rm E}^1\wedge f^1 +{\rm E}^2\wedge f^2 -2i ({\cal E}^1\wedge \bar{{\cal E}}^1+ {\cal E}^2\wedge \bar{{\cal E}}^2)\; ,
\ee
and
\be
{\rm d}\left({\rm d}\theta^\alpha \bar{\theta}_\alpha  -\theta^\alpha {\rm d}\bar{\theta}_\alpha \right)= -2 {\cal E}^1\wedge \bar{{\cal E}}^2 +2 {\cal E}^2\wedge \bar{{\cal E}}^1 \;  \qquad
\ee
are also useful to search for $\kappa-$symmetry of the single D$0$-brane in 3D. The formal exterior derivative of the Lagrangian 1-form of this system reads
\bea
{\rm d}{\cal L}^{\rm 3d\; \rm D0}&=& m {\rm d}\left({\rm E}^0 +{\rm d}\theta^\alpha \bar{\theta}_\alpha  -\theta^\alpha {\rm d}\bar{\theta}_\alpha \right) \nonumber \\ &=& -2im ({\cal E}^1 + i {\cal E}^2) \wedge (\bar{{\cal E}}^1-i\bar{{\cal E}}^2 )+ m{\rm E}^1\wedge f^1 +m{\rm E}^2\wedge f^2\nonumber \\ &=& -2im ({\cal E}^2 - i {\cal E}^1) \wedge (\bar{{\cal E}}^2+i\bar{{\cal E}}^1 )+ m{\rm E}^1\wedge f^1 +m{\rm E}^2\wedge f^2 \; .
\eea

\end{document}